\documentclass[12pt]{article}
\usepackage{epsfig,amsmath,amsfonts,amssymb,amstext,afterpage,psfrag,slashbox}
\long\def\symbolfootnote[#1]#2{\begingroup%
\def\thefootnote{\fnsymbol{footnote}}\footnote[#1]{#2}\endgroup}

\def\spose#1{\hbox to 0pt{#1\hss}}
\def\lsim{\mathrel{\spose{\lower 3pt\hbox{$\mathchar"218$}}
 \raise 2.0pt\hbox{$\mathchar"13C$}}}
\def\gsim{\mathrel{\spose{\lower 3pt\hbox{$\mathchar"218$}}
 \raise 2.0pt\hbox{$\mathchar"13E$}}}

\catcode`@=11
\def\@citex[#1]#2{%
  \if@filesw\immediate\write\@auxout{\string\citation{#2}}\fi
  \def\@citea{}\@cite{\@for\@citeb:=#2\do
    {\@citea\def\@citea{,\penalty\@m}\@ifundefined
      {b@\@citeb}{{\bf ?}\@warning
{Citation `\@citeb' on page \thepage \space undefined}}%
      \hbox{\csname b@\@citeb\endcsname}}}{#1}}
\def\citer{\@ifnextchar [{\@tempswatrue\@citexr}{\@tempswafalse\@citexr[]}}
  \def\@citexr[#1]#2{%
    \if@filesw\immediate\write\@auxout{\string\citation{#2}}\fi
    \def\@citea{}\@cite{\@for\@citeb:=#2\do
      {\@citea\def\@citea{--\penalty\@m}\@ifundefined
{b@\@citeb}{{\bf ?}\@warning
{Citation `\@citeb' on page \thepage \space undefined}}%
\hbox{\csname b@\@citeb\endcsname}}}{#1}}

\begin{document}

\begin{titlepage}

\begin{flushright}
{\small
CERN-PH-TH/2010-305\\
LMU-ASC~107/10\\ 
TUM-HEP-788/10\\
IPPP/11/02\\
DCPT/11/04\\
January 2011
%Draft \today
%hep-ph/yymmnnn
}
\end{flushright}

\vspace{0.5cm}
\begin{center}
{\Large\bf \boldmath
Theory of $B\to K^{(*)}l^+l^-$ decays at high $q^2$:\\
\vspace*{0.3cm}
OPE and quark-hadron duality
\unboldmath}
\end{center}

\vspace{0.5cm}
\begin{center}
{\sc M. Beylich$^2$, G. Buchalla$^{1,2}$ and 
Th. Feldmann$^3$\symbolfootnote[4]{Address after January~2011: \\
IPPP, Department of Physics, University of Durham, Durham DH1 3LE, UK}}
\end{center}

\vspace*{0.4cm}

\begin{center}
$^1$CERN, Theory Division, CH--1211 Geneva 23, Switzerland\\
\vspace*{0.2cm}
$^2$Ludwig-Maximilians-Universit\"at M\"unchen, Fakult\"at f\"ur Physik,\\
Arnold Sommerfeld Center for Theoretical Physics, 
D--80333 M\"unchen, Germany\\
\vspace*{0.2cm}
$^3$Physik Department, Technische Universit\"at M\"unchen,\\
James-Franck-Stra\ss e, D--85748 Garching, Germany
\end{center}

\vspace{1.5cm}
\begin{abstract}
\vspace{0.2cm}\noindent
We develop a systematic framework for exclusive rare $B$ decays of
the type $B\to K^{(*)}l^+l^-$ at large dilepton invariant mass $q^2$.
It is based on an operator product expansion (OPE) for the required 
matrix elements of the nonleptonic
weak Hamiltonian in this kinematic regime. Our treatment differs from
previous work by a simplified operator basis, the explicit calculation
of matrix elements of subleading operators, and by a quantitative 
estimate of duality violation. The latter point is discussed in detail,
including the connection with the existence of an OPE and an illustration
within a simple toy model.

\end{abstract}

\vspace*{2.5cm}

\vfill
\end{titlepage}

%%%%%%%%%%%%%%%%%%%%%%%%%%%%%%%%%%%%%%%%%%%%%%%%%%%%%%%%%%%%%%%%%
%   Introduction
%%%%%%%%%%%%%%%%%%%%%%%%%%%%%%%%%%%%%%%%%%%%%%%%%%%%%%%%%%%%%%%%%
\section{Introduction}
\label{sec:intro}

The rare decays $B\to K^{(*)}l^+l^-$ are among the most important probes
of flavour physics. They are potentially sensitive to dynamics beyond
the Standard Model (SM) and have been intensely studied in the 
literature \cite{Buchalla:2008jp}. 
Measurements have been performed at the $B$-meson factories
%\cite{Ishikawa:2003cp,Aubert:2006vb,Aubert:2008ju,Aubert:2008ps,Wei:2009zv}
\citer{Ishikawa:2003cp,Wei:2009zv}
and at the Fermilab Tevatron \cite{Pueschel:2010rm}.
Excellent future prospects for detailed
measurements are provided by the LHC experiments ATLAS, CMS, and LHCb at CERN
\cite{Buchalla:2008jp}, and, in the longer run, by Super Flavour Factories 
based on $e^+e^-$ colliders 
%\cite{Bona:2007qt,Kageyama:2006zd,Browder:2007gg,Browder:2008em}.
\citer{Bona:2007qt,Browder:2008em}.

The calculability of $B\to K^{(*)}l^+l^-$ decay rates and distributions 
benefits from the fact that these processes are, to first approximation, 
semileptonic modes. Correspondingly, the hadronic physics is described by 
$B\to K^{(*)}$ form factors, which multiply a perturbatively calculable 
amplitude. This simple picture is not exact because also the nonleptonic 
weak Hamiltonian at scale $\sim m_b$ has $B\to K^{(*)}l^+l^-$ matrix 
elements. The prominent example is given by hadronic
interactions of the form $(\bar sb)(\bar cc)$, where the charm quarks 
annihilate into $l^+l^-$ through a virtual photon. Such charm-loop 
contributions are more complicated theoretically than the form-factor terms. 
Even though the charm loops are subdominant numerically in the kinematical 
regions of interest, they cannot be completely neglected. In particular, 
the related uncertainty needs to be properly estimated in order to obtain 
accurate predictions.

We need to distinguish three regions in the dilepton invariant mass $q^2$, 
for which the properties of charm loops are markedly different. 
For $7\,{\rm GeV}^2\lsim q^2\lsim 15\,{\rm GeV}^2$ the presence of very narrow
$c\bar c$ resonances leads to huge violations of quark hadron duality
\cite{Beneke:2009az} and the hadronic backgrounds from $B\to K^{(*)}\psi$,
followed by $\psi\to l^+l^-$, dominate the short distance rate by two orders 
of magnitude. This region in $q^2$ can be removed by experimental cuts.
 
For $q^2\lsim 7\,{\rm GeV}^2$ the kaon is very energetic and the charm 
loops can be computed systematically in the heavy-quark limit using 
QCD factorization for $B$ decays into light-like mesons 
\cite{Beneke:1999br,Beneke:2000ry}. 
This approach was first employed for $B\to K^{(*)}l^+l^-$ in
\cite{Beneke:2001at}. The results have many applications. A summary with 
detailed references can be found in \cite{Buchalla:2008jp} (see also 
\cite{Khodjamirian:2010vf} for a recent analysis).

The high-$q^2$ region, $q^2\gsim 15\,{\rm GeV}^2$ has received
comparatively little attention. In this case the kaon energy is around
a ${\rm GeV}$ or below, and (soft-collinear) QCD factorization is less
justified, becoming invalid close to the endpoint of the spectrum
at $q^2=(m_B-m_K)^2$. On the other hand, the large value of $q^2$
defines a hard scale for the hadronic contribution to $B\to K^{(*)}l^+l^-$.
Consequently an operator product expansion (OPE) can be constructed, which
generates an expansion of the amplitude in powers of $E_K/\sqrt{q^2}$
(or $\Lambda_{QCD}/\sqrt{q^2}$). Charm loops, and other hadronic 
contributions, are thus approximated as effective interactions that are 
local on the soft scales set by $E_K$ and $\Lambda_{QCD}$. This simplifies
the computation substantially. In fact, to leading order in the OPE the
hadronic contribution reduces to a standard form-factor term. This picture 
has been first discussed at lowest order in the OPE in 
\cite{Buchalla:1998mt}, where it was applied to the endpoint region of 
$B\to K^{(*)}l^+l^-$ and $B\to K\pi l^+l^-$. In \cite{Grinstein:2004vb}
the OPE was considered in some detail, including a discussion of power
corrections.

In the present paper we formulate the OPE for the high-$q^2$ region
of $B\to K^{(*)}l^+l^-$ from the outset. Although our approach is similar in
spirit to the analysis of \cite{Grinstein:2004vb}, the concrete 
implementation is different. We will also go beyond the estimates
presented in \cite{Grinstein:2004vb} in several ways.
An important difference is that \cite{Grinstein:2004vb} combines the
OPE with heavy-quark effective theory (HQET), whereas we prefer
to work with $b$-quark fields in full QCD. The latter formulation
has the advantage of a simplified operator basis, which makes
the structure of power corrections and their evaluation considerably more
transparent. We also retain the kinematical dependence on $q^2$ in the 
coefficient functions, rather than expanding it around $q^2=m^2_b$.  
We further discuss the issue of quark-hadron duality, which appears
relevant because of the existence of $c\bar c$ resonance structure
in the $q^2$ region of interest. Violations of duality are effects 
beyond any finite order in the OPE. Using a resonance model based on
a proposal by Shifman, we quantify for the first time the size
of duality violations in the high-$q^2$ region of $B\to K^{(*)}l^+l^-$. 
Further aspects and new results of our analysis will be summarized in 
sections 8 and 9. The main conclusion is that $B\to K^{(*)}l^+l^-$
is under very good theoretical control also for $q^2\gsim 15\,{\rm GeV}^2$.
Precise predictions can be obtained in terms of the standard 
form factors, with essentially negligible effects from the additional
hadronic parameters related to power corrections and duality violation.

The paper is organized as follows.
Section 2 collects basic expressions for later reference.
In section 3 our formulation of the OPE for $B\to K^{(*)}l^+l^-$
at high $q^2$ is described and the power expansion is constructed explicitly,
complete to second order in $1/\sqrt{q^2}$ and with a discussion of weak 
annihilation as an example of a (small) third-order correction.
In section 4 we present an estimate of the matrix elements of 
higher-dimensional operators and quantify their impact on the decay 
amplitudes for both $B\to K$ and $B\to K^*$ transitions.
Section 5 discusses the connection between the OPE for large $q^2$
and QCD factorization for energetic kaons, which are shown to
give consistent results at intermediate $q^2\approx 15\,{\rm GeV}^2$.
In section 6 we address the subject of duality violation 
in the context of a toy model analysis. The estimate is then
adapted to the case of $B\to Kl^+l^-$ in section 7. In this
section we also address conceptual aspects relevant for the 
existence of the OPE and the notion of quark-hadron duality.   
A comparison of our approach with the literature is given in 
section 8 before we conclude in section 9.
Details on the basis of operators in the OPE are described
in appendix A and some numerical input is collected in appendix B.

%\afterpage{\clearpage}

%%%%%%%%%%%%%%%%%%%%%%%%%%%%%%%%%%%%%%%%%%%%%%%%%%%%%%%%%%%%%%%%%
%     Basic formulas
%%%%%%%%%%%%%%%%%%%%%%%%%%%%%%%%%%%%%%%%%%%%%%%%%%%%%%%%%%%%%%%%%
\section{Basic formulas}
\label{sec:basicform}

\subsection{Weak Hamiltonian}

The effective Hamiltonian for $b\to sl^+l^-$ transitions reads
\cite{Buchalla:1995vs,Buras:1994dj,Misiak:1992bc}
\begin{equation}\label{heff}
{\cal H}_{\rm eff}=\frac{G_F}{\sqrt{2}}\sum_{p=u,c}\lambda_p
\left[ C_1 Q^p_1 + C_2 Q^p_2 +\sum_{i=3,\ldots ,10} C_i Q_i\right]
\end{equation}
where
\begin{equation}\label{lamps}
\lambda_p=V^*_{ps}V_{pb}
\end{equation}
The operators are given by
\begin{equation}
\begin{aligned}
   Q_1^p &= (\bar p b)_{V-A} (\bar s p)_{V-A} \,, \\
   Q_3 &= (\bar s b)_{V-A} \sum{}_{\!q}\,(\bar q q)_{V-A} \,, \\
   Q_5 &= (\bar s b)_{V-A} \sum{}_{\!q}\,(\bar q q)_{V+A} \,, \\
   Q_7 &= \frac{e}{8\pi^2}\,m_b\,
    \bar s\sigma_{\mu\nu}(1+\gamma_5) F^{\mu\nu} b \,, \\
   Q_9 &= \frac{\alpha}{2\pi} (\bar sb)_{V-A} (\bar ll)_{V} \,,
\end{aligned}
\qquad\quad
\begin{aligned}
   Q^p_2 &= (\bar p_i b_j)_{V-A} (\bar s_j p_i)_{V-A} \,, \\
   Q_4 &= (\bar s_i b_j)_{V-A} \sum{}_{\!q}\,(\bar q_j q_i)_{V-A} \,, \\
   Q_6 &= (\bar s_i b_j)_{V-A} \sum{}_{\!q}\,(\bar q_j q_i)_{V+A} \,, \\
   Q_8 &= \frac{g}{8\pi^2}\,m_b\,
    \bar s\sigma_{\mu\nu}(1+\gamma_5) G^{\mu\nu} b \,, \\
   Q_{10} &= \frac{\alpha}{2\pi} (\bar sb)_{V-A} (\bar ll)_{A}
\end{aligned}
\label{qqi}
\end{equation}
Note that the numbering of $Q^p_{1,2}$
is reversed with respect to the convention of \cite{Buchalla:1995vs}.
Our coefficients $C_{9,10}$ correspond to $\tilde C_{9,10}$
in \cite{Buchalla:1995vs} and we include the factor of $\alpha/(2\pi)$
in the definition of $Q_{9,10}$.  
The sign conventions for the electromagnetic and
strong couplings correspond to the covariant derivative
$D_\mu=\partial_\mu +ie Q_f A_\mu + i g T^a A^a_\mu$. With these
definitions the coefficients $C_{7,8}$ are negative in the
Standard Model.

\subsection{Dilepton-mass spectra and short-distance coefficients}

We define the kinematic quantities $s=q^2/m^2_B$
(where $q^2$ is the dilepton invariant mass squared), $r_K=m^2_K/m^2_B$,
and
\begin{equation}\label{lkdef}
\lambda_K(s)=1+r^2_K+s^2-2 r_K-2s-2 r_K s
\end{equation}
The differential branching fractions for $\bar B\to\bar Kl^+l^-$ can 
then be written as \cite{Bartsch:2009qp}
\begin{equation}\label{dbkllds}
\frac{dB(\bar B\to\bar Kl^+l^-)}{ds} =
\tau_B\frac{G^2_F\alpha^2m^5_B}{1536\pi^5}|V_{ts}V_{tb}|^2\cdot 
 \lambda^{3/2}_K(s) f^2_+(s)\left(|a_9(Kll)|^2 +|a_{10}(Kll)|^2\right)
\end{equation}
The coefficient $a_9(Kll)$ contains the Wilson
coefficient $C_9(\mu)$ combined with the short-distance
parts of the $\bar B\to\bar Kl^+l^-$ matrix elements of operators 
$Q_1,\ldots, Q_8$. The coefficient $a_9(Kll)$
multiplies the matrix element of the local operator $Q_9$ in the decay
amplitude. The coefficient $a_{10}(Kll)=C_{10}$ of the operator $Q_{10}$ 
is determined by very short distances $\sim 1/M_W$ and is precisely known.

The corresponding formulas for $\bar B\to\bar K^* l^+l^-$ can for instance
be found in \cite{Buchalla:2000sk}.

%%%%%%%%%%%%%%%%%%%%%%%%%%%%%%%%%%%%%%%%%%%%%%%%%%%%%%%%%%%%%%%%%
%     B-> Mll at high q2
%%%%%%%%%%%%%%%%%%%%%%%%%%%%%%%%%%%%%%%%%%%%%%%%%%%%%%%%%%%%%%%%%
\section{\boldmath  OPE for $B\to M l^+l^-$ amplitudes at high $q^2$}
\label{sec:bmllhq}

\subsection{General structure}

The amplitudes for the exclusive decays $B\to M l^+l^-$,
where $M=K$, $K^*$, or a similar meson, are given by the matrix 
element of the effective Hamiltonian in (\ref{heff}) between the initial
$B$ meson and the  $M l^+l^-$ final state.  
The dominant contribution comes from the semileptonic operators
$Q_{9,10}$. Their matrix elements are simple in the sense that 
all hadronic physics is described by a set of $B\to M$ transition
form factors. This is also true for the electromagnetic operator $Q_7$.
The matrix elements of the hadronic operators $Q_1,\ldots, Q_6,Q_8$
are more complicated. They are induced by photon exchange and can be 
expressed through the matrix element of a correlator between the 
hadronic part of the effective Hamiltonian 
\begin{equation}\label{hphad}
H^p\equiv C_1 Q^p_1 + C_2 Q^p_2 + \sum_{i=3}^{6,8} C_i Q_i
\end{equation}
and the electromagnetic current of the quarks
\begin{equation}\label{jemq}
j^\mu\equiv Q_q\, \bar q\gamma^\mu q
\end{equation}
where $Q_q$ is the electric charge quantum number of quark flavour $q$
and a summation over $q$ is understood.
The decay amplitude may thus be written as
\begin{equation}\label{abmll}
A(\bar B\to\bar M l^+l^-)=-\frac{G_F}{\sqrt{2}}\frac{\alpha}{2\pi}\lambda_t
\left[\left(A^\mu_9+\frac{\lambda_u}{\lambda_t}A^\mu_{cu}\right)
\bar u\gamma_\mu v + A^\mu_{10}\, \bar u\gamma_\mu\gamma_5 v\right]
\end{equation}
where $\bar u$ and $v$ are the lepton spinors and
\begin{eqnarray}\label{a9cu10}
A^\mu_9 &=& C_9\, \langle\bar M|\bar s\gamma^\mu(1-\gamma_5)b|\bar B\rangle
-\frac{8\pi^2}{q^2} i\int d^4x\, e^{iq\cdot x}\,
\langle\bar M| T\, j^\mu(x) H^c(0)|\bar B\rangle \nonumber\\
&& +\, C_7\, \frac{2i m_b}{q^2} q_\lambda\, 
\langle\bar M|\bar s\sigma^{\lambda\mu}(1+\gamma_5)b|\bar B\rangle \nonumber\\
A^\mu_{cu} &=&
\frac{8\pi^2}{q^2} i\int d^4x\, e^{iq\cdot x}\,
\langle\bar M| T\, j^\mu(x) (H^u(0)-H^c(0))|\bar B\rangle \nonumber\\
A^\mu_{10} &=& 
C_{10}\, \langle\bar M|\bar s\gamma^\mu(1-\gamma_5)b|\bar B\rangle
\end{eqnarray}
For $b\to s$ transitions the contribution from $A^\mu_{cu}$
is suppressed by the prefactor $\lambda_u/\lambda_t$
and can be neglected.

Exploiting the presence of the large scale $q^2\sim m^2_b$, an operator
product expansion (OPE) can be performed for the non-local term
\begin{equation}\label{khdef}
{\cal K}^\mu_H(q)\equiv 
-\frac{8\pi^2}{q^2} i\int d^4x\, e^{iq\cdot x}\, T\, j^\mu(x) H^c(0)
\end{equation}
which describes the contribution of 4-quark operators to the
$b\to sl^+l^-$ amplitude. Such an OPE corresponds to integrating out 
the hard quark loop, leading to a series of local effective interactions 
for the high-$q^2$ region. To leading order in the large-$q^2$ expansion
this has been presented in \cite{Buchalla:1998mt}. A discussion
of the OPE including higher-order contributions has been given
in \cite{Grinstein:2004vb}.

Before going into more detail we discuss the basic structure of the OPE
for ${\cal K}^\mu_H$. The expansion may be written as
\begin{equation}\label{khco}
{\cal K}^\mu_H(q) = \sum_{d,n} C_{d,n}(q)\, {\cal O}^\mu_{d,n}
\end{equation}
The operators ${\cal O}_{d,n}$ are composed of quark and gluon fields and
have the flavour quantum numbers of $(\bar sb)$. They are ordered according 
to their dimension $d$ and carry an index $n$ labeling different operators
with the same dimension.  
The $C_{d,n}(q)$ are the corresponding Wilson coefficients, which can be
computed in perturbation theory. The large scales justifying the expansion
are $m^2_b$ and $q^2$. They are counted as quantities of the same order.
The coefficients then scale as $C_{d,n}\sim m^{3-d}_b$ in the heavy-quark 
limit. Since the matrix elements 
$\langle \bar M|{\cal O}_{d,n}|\bar B\rangle$ scale as $\sqrt{m_b}$, the
matrix element of each term in (\ref{khco}) behaves as $m^{7/2 - d}_b$.
Current conservation implies that all operators are transverse in $q$,
\begin{equation}\label{odntrans}
q_\mu {\cal O}^\mu_{d,n}\equiv 0
\end{equation}

It is convenient to work with the $b$-quark field in full QCD.
This field could be further expanded within heavy-quark effective theory
(HQET), in order to make the $m_b$-dependence fully explicit. In such an 
approach many additional operators would arise whose hadronic matrix 
elements are not readily known. In contrast, the advantage of using the
$b$-field in full QCD is that fewer operators appear and that the matrix 
elements of the leading ones are given by common form factors. In this method
the OPE becomes particularly transparent and we will adopt it here.

\begin{figure}[t]
\begin{center}
%\psfrag{x}[t]{$s$}                                                             
%\psfrag{y}[b]{$(dB/ds)/B$}                                                        
\resizebox{8cm}{!}{\includegraphics{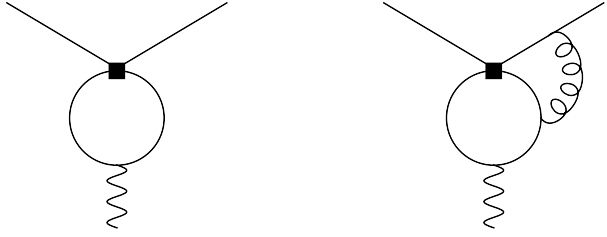}}
\caption{\label{fig:kh3}                                                       
OPE for the correlation function ${\cal K}^\mu_H$: Leading-power                
contributions (operators of dimension 3).                                       
The solid square and the virtual-photon attachment indicate the insertion of    
the weak Hamiltonian and of the electromagnetic current, respectively.          
The diagram on the left shows                                                   
the lowest-order term in QCD.  On the right is a sample diagram for             
the next-to-leading QCD corrections of order $\alpha_s$.}
\end{center}
\end{figure}

At leading order in the OPE ($d=3$), illustrated in Fig.~\ref{fig:kh3}, 
and in the chiral limit ($m_s=0$) there are two operators
\begin{eqnarray}
{\cal O}^\mu_{3,1} &=& \left(g^{\mu\nu}-\frac{q^\mu q^\nu}{q^2}\right)
\, \bar s\gamma_\nu(1-\gamma_5)b \label{o31} \\
{\cal O}^\mu_{3,2} &=& \frac{i m_b}{q^2} q_\lambda\,
\bar s\sigma^{\lambda\mu}(1+\gamma_5)b \label{o32}
\end{eqnarray}
Using the equations of motion for the external quarks it can be shown that
all possible bilinears $\bar s_L \Gamma b$ and 
$\bar s {\overleftarrow D}\Gamma b$ arising from the correlator
${\cal K}^\mu_H$ can be expressed in terms of (\ref{o31}) and (\ref{o32}).
Consequently, no independent dimension-4 operators of the form 
$\bar s {\overleftarrow D}\Gamma b$ can appear in the OPE. 
The complete proof is given in appendix A.
As an example, the operator $\bar s i\overleftarrow{D}^\mu(1+\gamma_5)b$
satisfies the equations-of-motion identity (for $m_s=0$)
\begin{equation}\label{sdb}
\bar s i\overleftarrow{D}^\mu(1+\gamma_5)b\equiv
-\frac{m_b}{2}\bar s\gamma^\mu(1-\gamma_5)b +
\frac{1}{2}\partial_\nu(\bar s\sigma^{\mu\nu}(1+\gamma_5)b)+
\frac{i}{2}\partial^\mu (\bar s(1+\gamma_5)b)
\end{equation}
For any $\bar B\to X_s$ matrix element with momentum transfer $q$ this is
equivalent to
\begin{equation}\label{sdbq} 
\bar s i\overleftarrow{D}^\mu(1+\gamma_5)b = 
-\frac{m_b}{2}\bar s\gamma^\mu(1-\gamma_5)b -
\frac{i}{2} q_\nu \bar s\sigma^{\mu\nu}(1+\gamma_5)b+
\frac{1}{2} q^\mu \bar s(1+\gamma_5)b                                  
\end{equation}
Because of current conservation only the transverse part
$(g_{\mu\lambda}-q_\mu q_\lambda/q^2)O^\lambda$ of such an operator $O^\mu$ can
appear in the OPE. From (\ref{sdbq}) we see that this part can be reduced to
a linear combination of (\ref{o31}) and (\ref{o32}).

If we keep $m_s\not= 0$, two additional operators have to be considered
\begin{eqnarray}
{\cal O}^\mu_{4,1} &=& m_s \left(g^{\mu\nu}-\frac{q^\mu q^\nu}{q^2}\right)
\, \bar s\gamma_\nu(1+\gamma_5)b \label{o41} \\  
{\cal O}^\mu_{4,2} &=& \frac{i m_s m_b}{q^2} q_\lambda\, 
\bar s\sigma^{\lambda\mu}(1-\gamma_5)b \label{o42} 
\end{eqnarray}
Since $m_s/m_b$ is small, and numerically similar to $\Lambda/m_b$, we 
may formally count these as operators of dimension 4.
Because they are absent at order $\alpha_s^0$,
their impact will be suppressed to the level of 
$\alpha_s m_s/m_b\sim 0.5\%$, which is negligible.
Note that these operators do in any case not introduce new hadronic
form factors.

\begin{figure}[t]
\begin{center}
%\psfrag{x}[t]{$s$}                                                            
%\psfrag{y}[b]{$(dB/ds)/B$}                                                    
\resizebox{8cm}{!}{\includegraphics{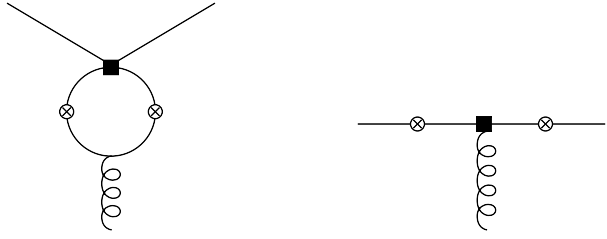}}
\caption{\label{fig:kh5}                                                      
OPE for the correlation function ${\cal K}^\mu_H$: Second-order power corrections 
(operators of dimension 5). The crossed circles denote the places where the      
virtual photon can be attached.}
\end{center}
\end{figure}

At $d=5$ (Fig.~\ref{fig:kh5})
we encounter operators with a factor of the gluon field strength
$G_{\mu\nu}$,  which have the form
\begin{equation}\label{o5n} 
{\cal O}^\mu_{5,n} = \bar s (gG\Gamma_n)^\mu b 
\end{equation}
where the $\Gamma_n$ denote Dirac and Lorentz structures.
We will treat the OPE explicitly to the level of $d=5$, that is including
power corrections up to second order in $\Lambda/m_b$. 

\begin{figure}[t]
\begin{center}
%\psfrag{x}[t]{$s$}                                                           
%\psfrag{y}[b]{$(dB/ds)/B$}                                                     
\resizebox{4cm}{!}{\includegraphics{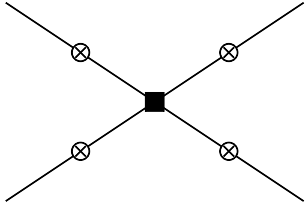}}
\caption{\label{fig:kh6}                                                       
OPE for the correlation function ${\cal K}^\mu_H$:                              
Weak annihilation as an example of third-order power corrections                
(operators of dimension 6). Crossed circles denote the various virtual           
photon attachments.}
\end{center}
\end{figure}

Although we will not give a full treatment of dimension-6 corrections,
we consider as an example the effect of weak annihilation 
(Fig.~\ref{fig:kh6}). 
This contribution
is characterized by the annihilation of the two valence quarks in the
$\bar B$ meson in the $\bar B\to \bar M$ transition through the
weak Hamiltonian. It is described by 4-quark operators, which read
schematically
\begin{equation}\label{o6ann}
{\cal O}^\mu_{6ann,n}=(\bar r\Gamma_1 b\, \bar s\Gamma_2 r)^\mu_n
\end{equation}
with Lorentz and Dirac structures indicated by $(\Gamma_{1,2})_n$, and
the light quark field $r=u$, $d$ in the case of non-strange $\bar B$
mesons. Weak annihilation provides a mechanism to break isospin
symmetry, directly at the level of the transition operator. 
In ${\cal K}^\mu_H$ weak annihilation, in addition to being a third order
power correction, comes only from QCD penguin operators, 
which have small coefficients. The contribution to isospin breaking from 
this source will therefore be strongly suppressed.

\subsection{\boldmath OPE to leading order in $\alpha_s$}

In this section we give explicitly the first few terms
in the OPE to leading order in renormalization-group improved perturbation
theory, that is neglecting relative corrections of ${\cal O}(\alpha_s)$.
This order for  ${\cal K}_H$ is relevant in the next-to-leading logarithmic
approximation to the $\bar B\to\bar M l^+l^-$ amplitude.    
We may then write
\begin{equation}\label{kh356}
{\cal K}^\mu_H = {\cal K}^\mu_{H3} +  {\cal K}^\mu_{H5} +
{\cal K}^\mu_{H6a} + {\cal O}(\alpha_s,(\Lambda/m_b)^3) 
\end{equation}
The lower indices of the terms on the r.h.s. denote the dimension $d$ of 
the corresponding local operators, which come with a coefficient of order
$1/m^{d-3}_b$. The first term reads
\begin{eqnarray}\label{kh3}
{\cal K}^\mu_{H3} &=& \left(g^{\mu\nu}-\frac{q^\mu q^\nu}{q^2}\right)
\bar s\gamma_\nu(1-\gamma_5) b \cdot \nonumber\\ 
&& \bigg[ h(z,\hat s) (C_1+3C_2+3C_3+C_4+3C_5+C_6)
- \frac{1}{2} h(1,\hat s) (4C_3+4C_4+3C_5+C_6) \nonumber\\
&& - \frac{1}{2} h(0,\hat s) (C_3+3C_4)+\frac{2}{9}(3C_3+C_4+3C_5+C_6)\bigg]
\end{eqnarray}
The coefficient in (\ref{kh3}) requires a UV renormalization, which has to
be consistent with the definition of $C_9$. The expression given here 
corresponds to the NDR scheme used in \cite{Buchalla:1995vs}.
The function $h(z,\hat s)$ is 
($z\equiv m_c/m_b$, $\hat s\equiv q^2/m^2_b$, 
$x\equiv 4z^2/\hat s=4m^2_c/q^2$)
\begin{equation}\label{hzs}
h(z,\hat s)=-\frac{8}{9}\ln\frac{m_b}{\mu}-\frac{8}{9}\ln z+\frac{8}{27}
+\frac{4}{9}x+\frac{2}{9}(2+x)\sqrt{1-x}
\left(\ln\frac{1-\sqrt{1-x}}{1+\sqrt{1-x}}+i\pi\right)
\end{equation}
Next we have
\begin{eqnarray}\label{kh5}
{\cal K}^\mu_{H5} &=& 
\bigg[\varepsilon^{\alpha\beta\lambda\rho}\frac{q_\beta q^\mu}{q^2} 
+\varepsilon^{\beta\mu\lambda\rho}\frac{q_\beta q^\alpha}{q^2}  
-\varepsilon^{\alpha\mu\lambda\rho} \bigg]\,
\bar s\gamma_\lambda(1-\gamma_5)gG_{\alpha\rho}b\,
\frac{C_1 Q_c}{q^2} f(x) \nonumber \\
&& -\frac{q_\lambda}{m_B} \bar s gG_{\alpha\beta}
(g^{\alpha\lambda}\sigma^{\beta\mu} - g^{\alpha\mu}\sigma^{\beta\lambda})
(1+\gamma_5)b\, \frac{4 C_8 Q_b}{q^2}
\end{eqnarray}
Here
\begin{equation}\label{fxdef}
f(x)=\frac{x}{\sqrt{1-x}}
\left(\ln\frac{1-\sqrt{1-x}}{1+\sqrt{1-x}}+i\pi\right) -2
\end{equation}
with $x=4m^2_c/q^2$.
The charm-loop contribution in (\ref{kh5}), proportional to $C_1$, 
can be inferred from \cite{Buchalla:1997ky}. 
Note that here we use the convention $\varepsilon^{0123}=-1$.
In writing (\ref{kh5}) we have neglected terms with the small
QCD penguin coefficients $C_3,\ldots, C_6$. 

Finally, weak-annihilation diagrams give the dimension-6
term
\begin{eqnarray}\label{kh6}
{\cal K}^\mu_{H6a} &=& \frac{8\pi^2}{q^4}q_\lambda\sum_{r=u,d} \bigg[ 2 Q_r 
(\bar r_i\gamma^\mu(1-\gamma_5)b_j\, \bar s_k\gamma^\lambda(1-\gamma_5)r_l
-\{ \mu\leftrightarrow\lambda\}) \nonumber\\
&& -\frac{2}{3}i\varepsilon^{\mu\lambda\beta\nu}  
\bar r_i\gamma_\beta(1-\gamma_5)b_j\, \bar s_k\gamma_\nu(1-\gamma_5)r_l\bigg]
\, (\delta_{ij}\delta_{kl} C_4 + \delta_{il}\delta_{kj} C_3)
\nonumber\\
&+&\frac{16\pi^2i}{q^4}q_\lambda\sum_{r=u,d} \bigg[ Q_r
(\bar r_i(1-\gamma_5)b_j\, \bar s_k\sigma^{\mu\lambda}(1+\gamma_5)r_l
+\bar r_i\sigma^{\mu\lambda}(1-\gamma_5)b_j\, \bar s_k(1+\gamma_5)r_l) 
\nonumber\\
&& -\frac{1}{3}
(\bar r_i(1-\gamma_5)b_j\, \bar s_k\sigma^{\mu\lambda}(1+\gamma_5)r_l
-\bar r_i\sigma^{\mu\lambda}(1-\gamma_5)b_j\, \bar s_k(1+\gamma_5)r_l)\bigg]
\nonumber\\
&& (\delta_{ij}\delta_{kl} C_6 + \delta_{il}\delta_{kj} C_5)
\end{eqnarray}
The terms in (\ref{kh6}) only arise from QCD penguin operators, which
have small coefficients.

We remark that all operators in (\ref{kh3}), (\ref{kh5}) and (\ref{kh6})
vanish identically when contracted with $q_\mu$, as required by
gauge invariance.

%%%%%%%%%%%%%%%%%%%%%%%%%%%%%%%%%%%%%%%%%%%%%%%%%%%%%%%%%%%%%%%%%%%%%%%%
\subsection{\boldmath
${\cal O}(\alpha_s)$ corrections to the charm loop}
\label{sec:alphas}

The non-factorizable ${\cal O}(\alpha_s)$ corrections to the charm loop
arise from
diagrams like the one shown on the r.h.s. of Fig.~\ref{fig:kh3}.
The $q^2$-dependence has been recently calculated in analytic form as a
Taylor
expansion in the small parameter $z=m_c^2/m_b^2$ \cite{Greub:2008cy}. 
Analytic results for $m_c=0$ had been presented in \cite{Seidel:2004jh}.
In the kinematical range
relevant to our considerations, it has been shown that the convergence
of the series is very
good. We therefore use the {\sc mathematica} input files provided by the
authors of \cite{Greub:2008cy} in
the online preprint publication for a numerical estimate. 
We find that the non-factorizable ${\cal O}(\alpha_s)$ corrections to
the charm loop
lead to a 10-15\% reduction of the real part of $a_9$ and contribute a
negative imaginary part 
of again 10-15\% relative to the short-distance contribution from $C_9$ 
(the precise value is scheme-dependent). 
This is in agreement with the effect found for the \emph{inclusive} $B
\to X_s l^+l^-$
rate in the high-$q^2$ region, as discussed in \cite{Greub:2008cy}, and
is similar to
the effect observed for the low-$q^2$ region in the exclusive decay
modes, see Table~5 in \cite{Beneke:2004dp}.

It is to be stressed that these corrections almost compensate 
the factorizable charm-loop contribution (diagram on the l.h.s. in
Fig.~\ref{fig:kh3}).
The reason why the  ${\cal O}(\alpha_s)$ corrections are not suppressed
stems from the different colour structure of the diagrams. Whereas the
factorizable
charm loop comes with a colour-suppressed combination of Wilson
coefficients, the additional
gluon exchange allows the $c\bar c$-pair to be in a colour-octet state
with no such suppression. 
At even higher orders in perturbation theory, ${\cal O}(\alpha^n_s)$
with $n\geq 2$, 
on the other hand, the numerical effect on $a_9$ should really be small,
as no new additionally
enhanced colour structures will arise.
%%%%%%%%%%%%%%%%%%%%%%%%%%%%%%%%%%%%%%%%%%%%%%%%%%%%%%%%%%%%%%%%%%

%%%%%%%%%%%%%%%%%%%%%%%%%%%%%%%%%%%%%%%%%%%%%%%%%%%%%%%%%%%%%               
%     Matrix elements 
%%%%%%%%%%%%%%%%%%%%%%%%%%%%%%%%%%%%%%%%%%%%%%%%%%%%%%%%%%%%%   
\section{Matrix elements and power corrections}
\label{sec:matrixel}

The computation of the amplitude from the OPE requires the
evaluation of the matrix elements of the local operators.
We estimate in particular the matrix elements of the
higher-dimensional contributions. This will allow us to   
quantify power corrections to the $B\to K^{(*)}l^+l^-$
amplitude at high $q^2$.
The cases of $B\to K$ and $B\to K^*$ transitions will be considered 
in turn. 

\subsection{\boldmath $B\to K$}
\label{subsec:btok}

The matrix element of the leading dimension-3 operator is given in terms
of the familiar form factors $f_\pm$, defined by ($p=k+q$)
\begin{equation}\label{bkffpm}
\langle \bar K(k)|\bar s\gamma^\mu(1-\gamma_5)b|\bar B(p)\rangle
=2 f_+(q^2)\, k^\mu + [f_+(q^2) + f_-(q^2)]\, q^\mu
\end{equation} 

At the level of the dimension-5 correction in (\ref{kh5})
one encounters operators of the form $\bar s G_{\alpha\rho}\Gamma b$.
Their matrix elements introduce, in general, new nonperturbative form
factors. Using Lorentz invariance and the antisymmetry of $G_{\alpha\beta}$
and $\sigma^{\rho\tau}$ one can show that 
\begin{equation}\label{bkkh5c8}
q_\lambda\, \langle \bar K(k)|\bar sG_{\alpha\beta}
(g^{\alpha\lambda}\sigma^{\beta\mu}-g^{\alpha\mu}\sigma^{\beta\lambda})
(1+\gamma_5)b|\bar B(p)\rangle\equiv 0
\end{equation}
In order to estimate the remaining term we assume 
$\Lambda\ll E_K\ll m_B$ for the kaon energy $E_K$. In this limit the 
matrix element can be computed in QCD factorization. 
To leading order we then find
\begin{equation}\label{bkkh5}
\langle \bar K(k)|{\cal K}^\mu_{H5}|\bar B(p)\rangle =
-\frac{\pi\alpha_s(E_K) C_F}{N} C_1 Q_c f(x)\frac{m_B f_B f_K}{\lambda_B\, q^2}
\, \left[k^\mu-\frac{k\cdot q}{q^2}q^\mu\right]
\end{equation}
where $C_F=(N^2-1)/(2N)$, $N$ is the number of colours, and $1/\lambda_B$
is the first inverse moment of the $B$-meson light-cone distribution
amplitude. This matrix element scales as $(\Lambda/m_b)^2$ relative
to (\ref{bkffpm}) in the heavy quark limit, and as ${\cal O}(1)$ for
large $N$.

In a similar way we can estimate the weak annihilation term
\begin{equation}\label{bkkh6}
\langle \bar K(k)|{\cal K}^\mu_{H6a}|\bar B(p)\rangle =
-\frac{16\pi^2 Q_r f_B f_K}{q^2}\,\left(C_4+\frac{C_3}{3}\right)
\, \left[k^\mu-\frac{k\cdot q}{q^2}q^\mu\right]
\end{equation}
This contribution is power-suppressed as $(\Lambda/m_b)^3$ relative
to (\ref{bkffpm}). 
The suppression by the small Wilson coefficients $C_{3,4}$ is partly
compensated by a large numerical factor of $\pi^2$.
To relative order $(\Lambda/m_b)^3$ there is no 
contribution from the term with $C_5$ and $C_6$.
Note that the result in (\ref{bkkh6}) also corresponds to the matrix
element obtained when naively factorizing the four-quark operators.

Normalized to the amplitude coefficient $a_9=C_9+\ldots$, the power 
corrections from ${\cal K}_{H5}$ and ${\cal K}_{H6a}$ read
\begin{equation}\label{dela9h5}
\Delta a_{9,H5}(K)= 
-\frac{\pi\alpha_s(E_K) C_F}{2N} C_1 Q_c f(x)\frac{m_Bf_Bf_K}{
    \lambda_B\, f_+(q^2)\, q^2}
\end{equation}
\begin{equation}\label{dela9h6}
\Delta a_{9,H6a}(K)= 
-\left(C_4+\frac{C_3}{3}\right)\,\frac{8\pi^2 Q_r f_B f_K}{f_+(q^2)\, q^2}
\end{equation}
where $r=u$, $d$ refers to the spectator quark in the $B$ meson.

Numerically, we find $\Delta a_{9,H5}(K)=0.019 - 0.012 i$ at
$q^2=15\,{\rm GeV}^2$ for central values of the parameters.
This number comes with a substantial uncertainty, in particular
from $\lambda_B$ and $f_+$. Nevertheless, the correction to 
$a_9\approx 4$ is very small, most likely below $1\%$ in magnitude. 
The correction in (\ref{dela9h5}) diminishes further for larger
$q^2$, reaching $\Delta a_{9,H5}(K)=0.006 - 0.002 i$ at the endpoint.
Towards the endpoint the kaon becomes soft and the result
in (\ref{dela9h5}), based on $E_K\gg\Lambda_{QCD}$, 
can only be viewed as a rough model calculation. The conclusion
that $\Delta a_{9,H5}(K)$ remains negligibly small should however still hold.
The weak annihilation correction is $\Delta a_{9,H6a}(K)=0.003$
at $q^2=15\,{\rm GeV}^2$ for $r=u$ and therefore entirely negligible, 
a consequence also of the small Wilson coefficients.

\subsection{\boldmath $B\to K^*$}
\label{subsec:btokst}

In the case of the decay into a vector meson the relevant form factors 
are defined as ($m_V=m_{K^*}$) 
\begin{flalign}
\big\langle \bar K^{*}(k,\varepsilon) \big| \bar s\gamma^\mu b \big|
\bar B (p)\big\rangle 
& = -2i \frac{V(q^2)}{m_B+m_V} \varepsilon^{\mu\nu\rho\sigma} 
\varepsilon_\nu^* p_\rho k_\sigma  \\
\big\langle\bar K^{*}(k,\varepsilon) \big| \bar s\gamma^\mu\gamma_5 b\big|
\bar B (p)\big\rangle 
& = 2 m_V A_0 (q^2) \frac{\varepsilon^* \cdot q}{q^2} q^\mu + 
(m_B + m_V) A_1(q^2) \bigg[{\varepsilon^*}^\mu-
\frac{\varepsilon^* \cdot q}{q^2}q^\mu \bigg] \nonumber \\
& - A_2(q^2)\frac{\varepsilon^* \cdot q}{m_B +m_V}\bigg[(p+k)^\mu - 
\frac{m_B^2 - m_V^2}{q^2} q^\mu\bigg]
\end{flalign}

It is convenient to treat the decay into longitudinally and transversely 
polarized vector mesons separately. Omitting terms proportional to $q^\mu$
\begin{eqnarray}\label{bkparff}
&&\langle \bar K^*_\parallel(k,\varepsilon)|
\bar s\gamma^\mu(1-\gamma_5)b|\bar B(p)\rangle
=\nonumber\\
&&\quad -2 k^\mu
\left[\frac{m_B+m_V}{2 m_V} A_1\frac{1-m^2_V/(m_B E)}{\sqrt{1-(m_V/E)^2}}
-\frac{m_B E \sqrt{1-(m_V/E)^2}}{m_V(m_B+m_V)} A_2\right]
\end{eqnarray}

\begin{equation}\label{bkperpff}
\langle \bar K^*_\perp(k,\varepsilon)|
\bar s\gamma^\mu(1-\gamma_5)b|\bar B(p)\rangle
=\frac{-2i V}{m_B+m_V}\varepsilon^{\mu\nu\rho\sigma}\varepsilon^*_{\perp\nu}
p_\rho k_\sigma -(m_B+m_V) A_1 \varepsilon^{*\mu}_\perp
\end{equation}

In the large energy limit $E\gg m_V$, which we may use in the 
normalization of the power corrections, (\ref{bkparff})
and (\ref{bkperpff}) simplify to \cite{Charles:1998dr,Beneke:2000wa}
\begin{eqnarray}\label{bkstappr}
\langle \bar K^*_\parallel(k,\varepsilon)|
\bar s\gamma^\mu(1-\gamma_5)b|\bar B(p)\rangle
&=&  -2 k^\mu A_0 \nonumber\\
\langle \bar K^*_\perp(k,\varepsilon)|
\bar s\gamma^\mu(1-\gamma_5)b|\bar B(p)\rangle &=&
-\frac{2 V}{m_B}\left(
i\varepsilon^{\mu\nu\rho\sigma}\varepsilon^*_{\perp\nu} p_\rho k_\sigma
+k\cdot p\, \varepsilon^{*\mu}_\perp\right)
\end{eqnarray}

The case of a longitudinally polarized $K^*$ is very similar
to the case of a pseudoscalar $K$, discussed in section \ref{subsec:btok},
and we find
\begin{equation}\label{dela9h5l}
\Delta a_{9,H5}(K^*_\parallel) =
-\frac{\pi\alpha_s(E_K) C_F}{2N} C_1 Q_c f(x)\frac{m_Bf_B f_\parallel}{
    \lambda_B\, A_0(q^2)\, q^2}
\end{equation}
where $f_\parallel$ is the decay constant of $K^*_\parallel$.

For a $K^*$ with transverse polarization we obtain
\begin{equation}\label{dela9h5t}
\Delta a_{9,H5}(K^*_\perp) =
-\frac{\pi\alpha_s(E_K) C_F}{4N} \frac{m_Bf_Bf_\perp}{
    \lambda_B\, V(q^2)\, q^2}
\left(C_1 Q_c f(x) + 8 C_8 Q_b\right)
\end{equation}
where $f_\perp$ is the decay constant of $K^*_\perp$.

Numerically we have
$\Delta a_{9,H5}(K^*_\parallel) = 0.021-0.012i$ and
$\Delta a_{9,H5}(K^*_\perp) =0.008-0.006i$ at $q^2=15\,\rm{GeV}^2$
for our standard set of parameters. These corrections are of similar
size as for the pseudoscalar kaon and they are likewise negligible.

Finally, we quote the corrections from weak annihilation
\begin{flalign}
\Delta a_{9,H6a}(K^*_\parallel) &= -\bigg(C_4+\frac{C_3}{3}\bigg)
\frac{8\pi^2 Q_r f_B f_\parallel}{A_0(q^2)q^2} \\[12pt]
\Delta a_{9,H6a}(K^*_\perp) &= -\bigg(C_6+\frac{C_5}{3}\bigg)
\frac{8\pi^2 f_B f_\perp m_B^2}{V(q^2)q^4} \bigg(Q_r-\frac{1}{3}\bigg)
\end{flalign}
where $r=u$, $d$ refers to the spectator quark in the B meson.
The corrections are tiny, at $q^2=15\,\rm{GeV}^2$ we have
$\Delta a_{9,H6a}(K^*_\parallel) = \Delta a_{9,H6a}(K^*_\perp) = 0.003$ 
for $r=u$, and 
$\Delta a_{9,H6a}(K^*_\parallel) = -0.001$,
$\Delta a_{9,H6a}(K^*_\perp) = -0.006$ for $r=d$.
This is again a consequence of the $1/q^2$ suppression   
at large values of $q^2$, see also \cite{Feldmann:2002iw}.

%%%%%%%%%%%%%%%%%%%%%%%%%%%%%%%%%%%%%%%%%%%%%%%%%%%%%%%%%%%%%%%%%
%     Large-small recoil                   
%%%%%%%%%%%%%%%%%%%%%%%%%%%%%%%%%%%%%%%%%%%%%%%%%%%%%%%%%%%%%%%%%
\section{Large vs. small recoil energy of the kaon}
\label{sec:largesmallrec}

The OPE for the correlator (\ref{khdef}), applied to
$B\to K^{(*)}l^+l^-$, is valid as long as the energy
of the kaon in the $B$-meson rest frame
\begin{equation}\label{ekaon}
E_K=\frac{m^2_B+m^2_K-q^2}{2 m_B}
\end{equation}
is small compared to $\sqrt{q^2}$. This condition is certainly
fulfilled in the vicinity of the endpoint, but even for 
$q^2$ as low as $15\,{\rm GeV}^2$, just above the narrow-resonance
region, $E_K=1.24\,{\rm GeV}$ is still fairly small in comparison
to the hard scale. On the other hand, such a value of $E_K$ 
is already larger than the QCD scale $\Lambda_{QCD}$ and one could
consider using the factorization methods applicable to the case of
energetic kaons. For $q^2$ at $15\,{\rm GeV}^2$, or somewhat above,
we have a transition region for the applicability of
factorization methods for large $E_K$ and the OPE for small $E_K$. 
It is this transition we wish to explore in the present section.

QCD factorization (QCDF) at large kaon recoil requires
$E_K\gg\Lambda_{QCD}$ for arbitrary $q^2$. The OPE method
requires $\sqrt{q^2}\gg E_K$, $\Lambda_{QCD}$.
Both scenarios are consistent with the case
\begin{equation}\label{q2eklambda}
\sqrt{q^2}\gg E_K\gg\Lambda_{QCD}
\end{equation}
which can be realized, at least approximately, as we have seen above.
In both, the QCDF and the OPE scenario, the leading contributions to the 
$B\to K^{(*)}l^+l^-$ amplitudes are given by short-distance
quantities multiplying the standard $B\to K^{(*)}$ form factors.
In QCDF, the first corrections to these form-factor terms come from
hard-spectator interactions, which have been computed in \cite{Beneke:2001at}. 
Normalized to $a_9$ (called ${\cal C}_9$ in \cite{Beneke:2001at})
these corrections may be written as 
$\Delta a^{(nf)}_{9,\parallel +} + \Delta a^{(nf)}_{9,\parallel -}$, 
where the indices refer to the notation of \cite{Beneke:2001at}.
For definiteness we will consider the case of a pseudoscalar kaon
in the following.
Neglecting the small penguin coefficients $C_3,\ldots, C_6$ and
adapting the expressions in \cite{Beneke:2001at} to our notation,
the second contribution reads ($\bar u\equiv 1-u$)
\begin{eqnarray}\label{da9m}
\Delta a^{(nf)}_{9,\parallel -} &=& -\frac{\pi\alpha_s C_F}{2N}
\frac{m_b f_B f_K Q_q}{m_B f_+(q^2)}
\int\frac{d\omega\, \phi_-(\omega)}{m_B\omega-q^2-i\epsilon}
\nonumber\\
&&\cdot\int_0^1 du\,\phi_K(u)\left[\frac{8C_8}{\bar u+u\frac{q^2}{m^2_B}}+
\frac{6 m_B}{m_b} C_1 h(z,\hat s)|_{q^2\to \bar u m^2_B + u q^2} \right]
\end{eqnarray}
In the OPE limit, defined by treating $q^2\sim m^2_B$ and taking
(\ref{q2eklambda}), we have $\Delta a^{(nf)}_{9,\parallel -}\sim 1/m^3_B$.
This term is therefore subleading with respect to (\ref{dela9h5}), consistent
with the absence of a $C_8$ term in $\Delta a_{9,H5}(K)$.  
The remaining term is given by
\begin{equation}\label{da9p}
\Delta a^{(nf)}_{9,\parallel +}=-\frac{\pi\alpha_s C_F}{2N} C_1 Q_c
\frac{f_B f_K}{m_B f_+(q^2)}\int\frac{d\omega}{\omega}\phi_+(\omega)
\int_0^1 du\,\phi_K(u) t_{\parallel}
\end{equation}
where ($E=E_K$)
\begin{equation}\label{tpardef}
t_\parallel=\frac{2 m_B}{\bar u E} I_1+
\frac{\bar u m^2_B+ u q^2}{\bar u^2 E^2}(B_0(\bar u m^2_B+u q^2)-B_0(q^2))
\end{equation}
\begin{equation}\label{b0def}
B_0(s)=\sqrt{1-\frac{4 m^2_c}{s}}\left(
\ln\frac{1-\sqrt{1-\frac{4 m^2_c}{s}}}{1+\sqrt{1-\frac{4 m^2_c}{s}}}
+ i\pi \right)
\end{equation}
\begin{equation}\label{i1def}
I_1=1+\frac{2 m^2_c}{\bar u(m^2_B-q^2)}(L_1(x_+)+L_1(x_-)-
L_1(y_+)-L_1(y_-))
\end{equation}
\begin{equation}\label{xydef}
x_\pm=\frac{1}{2}\pm\left(
\frac{1}{4}-\frac{m^2_c}{\bar u m^2_B+u q^2}\right)^{1/2}\pm i\epsilon\, ,
\qquad
y_\pm=\frac{1}{2}\pm\left(
\frac{1}{4}-\frac{m^2_c}{q^2}\right)^{1/2}\pm i\epsilon
\end{equation}
\begin{equation}\label{l1def}
L_1(w_\pm)={\rm Li}_2\left(\frac{w_\pm}{w_\pm-1}\right)-\frac{\pi^2}{6}+
\ln(1-w_\pm)\left(\ln\frac{1-w_\pm}{w_\pm}\pm i\pi\right)\, ,
\qquad w_\pm=x_\pm,\, y_\pm
\end{equation}
The formulas from \cite{Beneke:2001at} have been slightly rewritten
here to make them more convenient for application in the high-$q^2$
region where $4 m^2_c/q^2 < 1$.
To make contact with the OPE result we eliminate $q^2$ from (\ref{tpardef})
using $q^2=m^2_B-2 m_B E$ and expand in $E/m_B$. The terms proportional
to $1/E^2$ and $1/E$ cancel and we find
\begin{equation}\label{tparfx}
t_\parallel = f(x) + {\cal O}\left(\frac{E}{m_B}\right)
\end{equation}
with the function $f(x)$ from (\ref{fxdef}). 
In the limit $E_K\ll m_B$ the QCDF formula
(\ref{da9p}) then reduces to the OPE result (\ref{dela9h5}).
We note that the hard-spectator correction (\ref{da9p}),
which is a leading-power contribution in QCDF, becomes a
power-correction $\sim 1/q^2\sim 1/m^2_B$ in the OPE regime.
This behaviour is related to the form factor $f_+(q^2)$,
which scales as $1/m^{3/2}_B$ at small $q^2$ and as $m^{1/2}_B$ 
at large $q^2$. 

We may use the preceding comparison to check the validity of
the OPE at the relatively low values of $q^2$ around $15\,{\rm GeV}^2$.
Assuming that the large-energy limit for the kaon is a reasonable
approximation, the QCDF expression (\ref{da9p}) and the OPE result
(\ref{dela9h5}) differ only by the replacement  
\begin{equation}\label{da9ph5}
\int_0^1 du\,\phi_K(u) t_{\parallel}= -5.05 + 2.70 i \qquad 
\to\qquad  f(x)\frac{m^2_B}{q^2}=-5.87 + 3.55 i
\end{equation}
where the numerical values are obtained at $q^2=15\,{\rm GeV}^2$ 
for our standard set of parameters, neglecting the kaon mass for 
consistency. The real part of the OPE approximation
is $16\%$ larger in magnitude than the QCDF result, for the
imaginary part the discrepancy is about $30\%$. Such differences are
to be expected since the expansion leading from QCDF to the OPE is
governed by $E_K/\sqrt{q^2}\approx 0.3$. Within this accuracy, the OPE
formula still gives a very good estimate of the more complete QCDF
result at $q^2=15\,{\rm GeV}^2$. For larger values of $q^2$ the
numerical difference between the two sides in (\ref{da9ph5}) 
becomes smaller and towards the endpoint the OPE is the 
more appropriate description.
  
The above exercise suggests that even at $q^2$ as low as $15\,{\rm GeV}^2$
the OPE is a valid method to obtain the second-order correction
(\ref{dela9h5}). The difference with the QCDF estimate is immaterial
in view of the very small overall size of the effect.

The transition from large to small recoil energy has already been 
considered in \cite{Bartsch:2009qp} for the $B\to K$ form factor ratio 
$f_T/f_+$. Also in this and in analogous cases, hard-spectator corrections, 
which are leading-power effects at large recoil, become power suppressed 
for soft kaons. 
Similar observations hold for weak annihilation \cite{Bartsch:2009qp}.

%%%%%%%%%%%%%%%%%%%%%%%%%%%%%%%%%%%%%%%%%%%%%%%%%%%%%%%%%%%%%%%%%
%     Duality: toy model                   
%%%%%%%%%%%%%%%%%%%%%%%%%%%%%%%%%%%%%%%%%%%%%%%%%%%%%%%%%%%%%%%%%
\section{Toy-model analysis of duality violation}
\label{sec:dualitytoy}

The OPE defines a systematic framework to compute the
correlator (\ref{khdef}) at high $q^2$ in QCD. 
In the Minkowski region $q^2 >0$, there will be
uncertainties in the OPE-based predictions that
go beyond those due to neglected orders in $\alpha_s$
or in $\Lambda/m_b$. Such effects are refered to as
violations of quark-hadron duality. We investigate
their importance first within a toy model 
for the charm loops in $B\to Kl^+l^-$. 
In section 7 the model is generalized to be closer
to the realistic case.   

\subsection{Description of the model}
\label{subsec:toydesc}

It is illuminating to consider the systematics of duality
for the charm-loop contribution to $B\to Kl^+l^-$ 
in the simplified context of the toy model introduced
in \cite{Beneke:2009az}.
This model assumes the existence of two leptons, $l_1$ with a large mass $m_1$ 
and $l_2$ with mass $m_2=0$, and the effective weak Hamiltonian
\begin{equation}\label{htoy}
   {\cal H}_{\rm eff} 
   = \frac{G}{\sqrt{2}} \left[ 
   (\bar l_2l_1)_{V-A}\,(\bar cc)_{V-A} 
   - (\bar l_2l_1)_{V-A}\,(\bar tt)_{V-A} \right] 
\end{equation}
All particles have standard strong and electromagnetic interactions. 
Then ${\cal H}_{\rm eff}$ gives rise to a loop-induced process 
$l_1\to l_2\,e^+e^-$ via 
charm- and top-quark penguin diagrams with a GIM-like cancellation between 
them. The hadronic physics is fully contained in current correlators 
of the form
\begin{equation}\label{pimunu}  
   \Pi^{\mu\nu}_c 
  = i\int\!d^4x\,e^{iq\cdot x}\, 
\langle 0|T\,j^\mu_c(x) j^\nu_c(0)|0\rangle   
\equiv (q^\mu q^\nu - q^2 g^{\mu\nu})\,\Pi_c(q^2)
\end{equation}
with $j^\mu_c=\bar c\gamma^\mu c$. The decay amplitude can then be written as  
\begin{equation}\label{al12ee}
   A(l_1\to l_2\,e^+e^-)
   = - \frac{G}{\sqrt{2}}\,e_c e^2\,\Pi(q^2)\,
   \bar l_2\gamma^\mu(1-\gamma_5)l_1\,\bar e\gamma_\mu e 
\end{equation}
where $\Pi\equiv\Pi_c-\Pi_t$ is the difference of the charm and top 
contributions. We take $m_t > m_1$, and thus $ \mbox{Im}\,\Pi$ comes only
from the charm sector. The correlator $\Pi(q^2)$ fulfills the dispersion 
relation 
\begin{equation}\label{disprel} 
 \Pi(q^2) - \Pi(0) = \frac{q^2}{\pi} \int_0^\infty\!\frac{dt}{t}\,
 \frac{ \mbox{Im}\,\Pi(t)}{t-q^2-i\epsilon} 
\end{equation}
where the subtraction constant $\Pi(0)$ is fixed within the model and can 
be computed in perturbation theory. To leading order it reads
\begin{equation}\label{pi0}
   \Pi(0)\equiv \Pi_c(0)-\Pi_t(0)
   = \frac{N}{12\pi^2}\ln\frac{m^2_t}{m^2_c} 
\end{equation}
The form of (\ref{al12ee}) for the amplitude holds to lowest order in $G$ 
and $e^2$, but to all orders in the strong coupling. 
The decay $l_1\to l_2 e^+e^-$ in this model shares important similarities with 
$B\to Ke^+e^-$, but the hadronic dynamics is simplified to the physics
of quark-current correlators. The role of resonances and quark-hadron duality
can thus be illustrated in a transparent way.

From (\ref{al12ee}) we obtain the differential decay rate (with $s=q^2/m_1^2$)
\begin{equation}\label{gaml12ee}
   \frac{d\Gamma(l_1\to l_2\,e^+e^-)}{ds}
   = \frac{G^2\alpha^2m^5_1}{108\pi^5}\,(1-s)^2\,(1+2s)\,
   \big| C + \Delta(q^2) \big|^2 
\end{equation}
where we defined
\begin{equation}\label{ccdef}
C\equiv 2\pi^2\, \Pi(0)
\end{equation}
and
\begin{equation}\label{deldef} 
\Delta(q^2)\equiv 2\pi^2\, \left(\Pi(q^2)- \Pi(0)\right) 
\end{equation}
To lowest order we have
\begin{equation}\label{cclnmtmc}
 C = \ln\frac{m_t}{m_c} 
\end{equation}
and the partonic expression for $\Delta(q^2)$ in one-loop approximation 
is
\begin{equation}\label{delpart}                                                
   \Delta_q(q^2)                                                
   = \frac{5}{6} + \frac{x}{2}          
   - \frac{1}{4}\,(2+x) \sqrt{|1-x|}\, \left\{                                
   \begin{array}{lc}                                                          
   2\arctan\frac{1}{\sqrt{x-1}} \,; &  x>1  \\[3mm]                        
   \ln\frac{1+\sqrt{1-x}}{1-\sqrt{1-x}} - i\pi \,;~ & x<1                  
   \end{array}                                                                
   \right.                                                                    
\end{equation}
where $x=4m^2_c/q^2$. For typical values of the parameters 
(e.g.\ $m_c=1.4$\,GeV, 
$m_1=m_b=4.8$\,GeV, $m_t=167$\,GeV) we have $C=4.78$, whereas 
${\rm Re}\Delta_q(q^2)$ first rises from 0 at $q^2=0$ to 4/3 at
$q^2=4m^2_c$ and then drops again monotonically to the small negative value 
$-0.07$ at $q^2=m^2_b$. 
The coefficient $C$ represents the short-distance contribution of the
amplitude. It is real and larger (parametrically as well as numerically) than 
the quark-level charm contribution $|\Delta_q|$. 

The decay rate (\ref{gaml12ee}) is proportional to
\begin{equation}\label{ccdel}
|C+\Delta|^2=C^2 + 2 C\, {\rm Re}\Delta +|\Delta|^2
\end{equation}
As long as $|\Delta|$ is small compared to $C$, there is a
clear hierarchy among the three terms on the r.h.s. of (\ref{ccdel}):
The first, short-distance term $C^2$ dominates and the next term
gives the correction to first-order in $\Delta$, whereas the final term 
enters at second order. 

These features are qualitatively similar in the case of $B\to Kl^+l^-$.

\subsection{Shifman model for charm correlator}

In order to investigate the systematics of duality violation
in $l_1\to l_2 e^+e^-$, we find it convenient to consider first
a simple model for the quark-current correlator, which has 
been proposed in \cite{Blok:1997hs,Shifman:2000jv,Shifman:2003de}.
In this model the correlator is represented as an infinite sum over
resonances, which include finite width effects. In its original form
it applies to massless quarks and we will correspondingly neglect
the charm-quark mass in the present section.
A detailed discussion of the model and its use in illustrating
duality violation in the $R$ ratio and similar quantities
has been given in \cite{Blok:1997hs,Shifman:2000jv,Shifman:2003de}.
The model has also been used to study duality violation in 
$\tau$ decays in \cite{Cata:2008ye}.

In Shifman's model the correlator $\Delta$ in (\ref{deldef}) reads
\begin{equation}\label{delshm}
\Delta(q^2)=-\frac{N}{6}\frac{1}{1-b/\pi}
\left[\psi(z+1)+\gamma\right]
\end{equation}  
where $\psi(z)=d\ln\Gamma(z)/dz$ is the digamma function and
\begin{equation}\label{zdef}
z=(-r-i\epsilon)^{1-b/\pi}\qquad {\rm with} \quad r=\frac{q^2}{\lambda^2}
\end{equation} 
$N=3$ is the number of colours, $\lambda$ is a scale corresponding
to the string tension in QCD and $b\equiv B/N=\Gamma_n/M_n$ is a 
(small) parameter related to the width-to-mass ratio of the resonances. 

The model expression for $\Delta$ in (\ref{delshm}) has the correct
analytic behaviour (a cut for positive $q^2$ but no other singularities
on the physical sheet) and it reproduces the asymptotic result of QCD
in the limit of large $q^2$,
\begin{equation}\label{dellog}
\Delta(q^2) \to -\frac{N}{6}\ln\frac{-q^2-i\epsilon}{\lambda^2}
\end{equation}     

\begin{figure}[t]
\begin{center}
%\psfrag{x}[t]{$s$}
%\psfrag{y}[b]{$(dB/ds)/B$}
\resizebox{8cm}{!}{\includegraphics{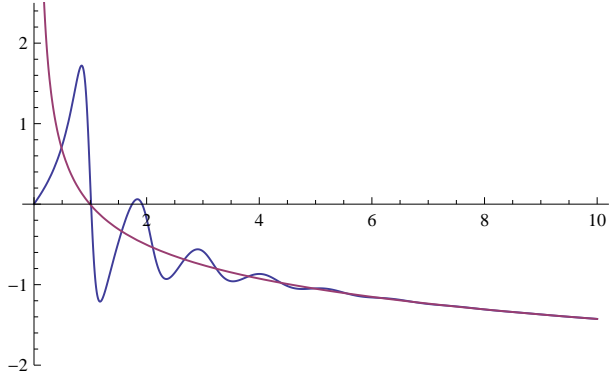}}
\caption{\label{fig:rddvope}
Shifman model for charm loop:
${\rm Re}\,\Delta(q^2)$ as a function of $q^2/\lambda^2$
for $b\equiv B/N=1/6$.
The true function (oscillating curve) is compared with
the OPE approximation.} 
\end{center}
\end{figure}

Using the identity
\begin{equation}\label{psiid}
\psi(z+1)+\gamma\equiv \left[\psi(-z)+\gamma-i\pi \right]_1
+ \left[-\pi \cot \pi z + i\pi \right]_2
\end{equation}
the function in (\ref{delshm}) can be decomposed into two parts,
$\Delta=\Delta_1+\Delta_2$, corresponding to the two brackets
in (\ref{psiid}). $\Delta_2$ is an oscillating function of $q^2$,
exponentially suppressed for large $q^2$.
It represents the duality violating component of $\Delta$. 
The function $\Delta_1$ is a monotonous function, which gives
the OPE approximation to $\Delta$. The real part
of these two functions is shown in Fig.~\ref{fig:rddvope}. 
A plot of the imaginary part of $\Delta$ can be found in
\cite{Shifman:2000jv,Shifman:2003de}.

The duality violating part of ${\rm Re}\Delta$ can be approximated as
\begin{equation}\label{deldv}
{\rm Re}\Delta_2(q^2)= -\frac{N}{6}\frac{1}{1-b/\pi}
                {\rm Re}\left[-\pi \cot \pi z + i\pi \right]\approx 
      -\frac{N\pi}{3} \exp(-2\pi b r) \sin(2\pi r)
\end{equation}     
if $2\pi b r\gg 1$ and $b\ln r/\pi \ll 1$.

\begin{figure}[t]
\begin{center}
%\psfrag{x}[t]{$s$}
%\psfrag{y}[b]{$(dB/ds)/B$}
\resizebox{8cm}{!}{\includegraphics{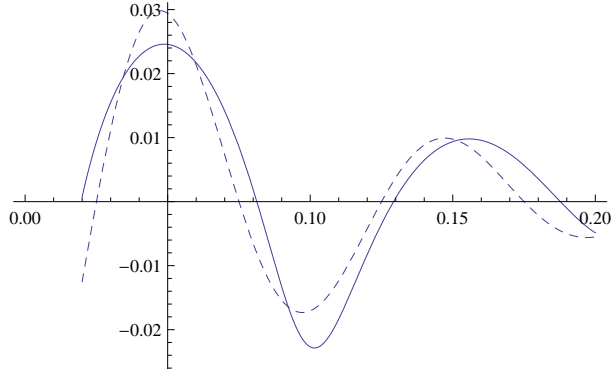}}
\caption{\label{fig:rddvint}
Shifman model for charm loop: Duality violating contribution
$\int^1_{s_0}ds\, (1+2s)(1-s)^2 {\rm Re}\,\Delta_2$ as a function of 
$s_0$ with parameters $b=1/6$ and $u=10$. 
The full result (solid curve) is compared with
the approximation given as the last term
in eq. (\ref{dvrate}) (dashed curve).}
\end{center}
\end{figure}

In the decay rate integrated over the high-$q^2$ part of the spectrum,
the duality violating contribution enters proportional to 
\begin{eqnarray}\label{dvrate}
&&\int^1_{s_0} ds\, (1+2s)(1-s)^2\, {\rm Re}\Delta_2
\approx -\frac{N\pi}{3}\int^1_{s_0} ds\, (1+2s)(1-s)^2\, 
\exp(-2\pi b u s) \sin(2\pi u s) \nonumber\\
&& = -\frac{N}{6}(1+2s_0) (1-s_0)^2\frac{1}{u}
\exp(-2\pi b s_0 u) \cos(2\pi s_0 u) + 
{\cal{O}}\left(\frac{b}{u},\frac{1}{u^2}\right)
\end{eqnarray}     
where
\begin{equation}\label{rus}
s=q^2/m^2_1 \qquad\quad   u=m^2_1/\lambda^2 \qquad\quad r= u s
\end{equation}
In (\ref{dvrate}) we have used the approximation from (\ref{deldv}).
Typical values of the parameters are
\begin{equation}\label{bupar}
b=\frac{1}{6}\qquad\quad  u=10 
\end{equation}
The value of $u=10$ corresponds for instance to 
$\lambda^2=2.3\,{\rm GeV}^2$ and $m^2_1=23\,{\rm GeV}^2$.
The quantity in (\ref{dvrate}) is shown in Fig.~\ref{fig:rddvint}
as a function of $s_0$. 

We comment on several important aspects of these results.
\begin{itemize}
\item
The duality violating component of ${\rm Re}\Delta$ in (\ref{deldv})
exhibits the characteristic oscillating behaviour in $r=q^2/\lambda^2$
with an exponential suppression governed by $br$.
The analogous expression for the duality violating term in 
${\rm Im}\Delta$, which has a cosine instead of the sine,
has been given in \cite{Shifman:2000jv,Shifman:2003de}.
\item
Eq. (\ref{dvrate}) displays the duality violating contribution from
${\rm Re}\Delta$ to the partially integrated decay rate.
The integration over $s$ extends from a suitably chosen lower limit
$s_0$ up to the end of the spectrum. The parameter $s_0$ should be large
enough such that the OPE still remains reasonable at the corresponding 
value of $q^2$. Using the approximation in (\ref{deldv}) and expanding in 
the small quantities $b$ and $1/u$, we find the explicit result written
as the last expression in (\ref{dvrate}). 
This expression is an oscillating function of $s_0$, with frequency $2\pi u$,
multiplied by an exponential suppression factor. The latter is
active when the exponent is large, at least $2\pi b s_0 u\sim 1$.
In addition, the entire term is further suppressed by one power of
$1/u=\lambda^2/m^2_1$. This effect remains even if the exponential 
suppression is not fully developped. The power suppression arises because
the oscillating contributions average out in the integral, except for 
a remainder $\sim 1/u$ near the lower end of integration $s_0$.
\item
The duality violating term (\ref{dvrate}) is plotted in 
Fig.~\ref{fig:rddvint} for a semi-realistic choice of parameters, 
where $m_1=4.8\,{\rm GeV}$ and $\lambda^2=2.3\,{\rm GeV}^2$.
The smallest values of $s_0$ shown, $s_0\sim 0.05$, correspond
to $q^2\sim 1.2\,{\rm GeV}^2$ and $2\pi b s_0 u\sim 0.5$, which is
already on the low side of the allowed range. 
From Fig.~\ref{fig:rddvint} we observe, first, that in the scenario
considered here the simple approximation to (\ref{dvrate}) agrees very
well with the full result. Second, the numerical size of the duality 
violating term is about $\pm 0.02$ for the relatively low
values of $s_0$ around $0.1$. The effect diminishes quickly for larger $s_0$
due to the exponential suppression.
The variation $\pm 0.02$ from (\ref{dvrate}) amounts to
$\pm 6.5\%$  when compared with the corresponding OPE expression 
\begin{equation}
\int^1_{0.1}ds\, (1+2s)(1-s)^2\, {\rm Re}\,\Delta_1=-0.31
\end{equation} 
\end{itemize}

We next turn to a discussion of the $|\Delta|^2$ term
in (\ref{ccdel}), where the situation is systematically different
from the case of ${\rm Re}\Delta$. The $|\Delta|^2$ term receives
a contribution from the duality violating component given by
\begin{eqnarray}\label{delsqdv}
&&\int^1_{s_0}ds\, (1+2s)(1-s)^2\, |\Delta_2|^2\approx
\int^1_{s_0}ds\, (1+2s)(1-s)^2\, \left(\frac{N\pi}{3}\right)^2
\exp(-4\pi b u s) \nonumber\\
&&=\left(\frac{N\pi}{3}\right)^2 \frac{(1+2s_0)(1-s_0)^2}{4\pi b u}
\exp(-4\pi b s_0 u) + {\cal{O}}\left(\frac{1}{(bu)^2}\right)
\end{eqnarray}
where in the second step an approximation similar to the one
in (\ref{deldv}) has been used.
There is still an exponential suppression, which makes
the entire term negligible for sufficiently large $b s_0 u$.
On the other hand, the power suppression with $1/u$ observed in
(\ref{dvrate}) is softened into a behaviour as $1/(bu)$.
For small $b$ (comparable to $1/u$ or smaller) the
violation of duality may become large. This is in qualitative
agreement with the discussion of duality violation for the squared
correlator $|\Pi(q^2)|^2$ given in \cite{Beneke:2009az}. 
The enhanced impact of duality violation is related to the
absence of oscillations with alternating sign in the integrand
of (\ref{delsqdv}).   
We conclude that the $|\Delta|^2$ term in the integrated rate
is particularly susceptible to violations of quark-hadron duality,
which may lead to substantial (positive) deviations from the OPE result,
unless $q^2$ is large enough for a strong exponential suppression
of the effect. However, the uncertainties in the
$|\Delta|^2$ term may be immaterial if this contribution
is only a small correction to the dominant $C^2$ part in (\ref{ccdel}).
 
Let us finally illustrate the relative importance of the various 
contributions to the rate of $l_1\to l_2 e^+e^-$ using the Shifman
model for the charm loop with the numerical input defined above. 
In the limit $m_c\to 0$ considered here the function $C$ from 
(\ref{cclnmtmc}) is modified to
\begin{equation}\label{ccmc0}
C=\ln\frac{m_t}{\lambda} + \frac{\gamma}{2} + \frac{5}{6} = 5.82
\end{equation}
The relative contributions to the decay rate (\ref{gaml12ee}), 
(\ref{ccdel}) then read
\begin{eqnarray}
\int^1_{0.1} ds\, g(s)\, C^2 &=& 13.60 \label{ccnum}\\
\int^1_{0.1} ds\, g(s)\, 2C\, {\rm Re}\Delta_1 
     &=& -3.59 \qquad [\pm 0.23]\label{cdnum}\\ 
\int^1_{0.1} ds\, g(s)\, |\Delta_1|^2 
     &=& 1.25 \qquad [+0.10]\label{ddnum}
\end{eqnarray}
with $g(s)=(1+2s)(1-s)^2$. The central values in (\ref{cdnum}) and
(\ref{ddnum}) are based on the OPE result for $\Delta$, the square brackets
indicate the impact of duality violation.
The uncertainty in (\ref{cdnum}) gives the variation due to
${\rm Re}\Delta_2$. It has the relative size of $\pm 6.5\%$, which
is reduced to $\pm 2\%$ in the sum of all contributions.
The shift from duality violation in (\ref{ddnum}) is positive and
essentially negligible in the present example.

\subsection{Model for charm correlator based on BES data}
\label{sec:charmbes}

\begin{figure}[t]
 \begin{center}
  \includegraphics[width=0.45\textwidth]{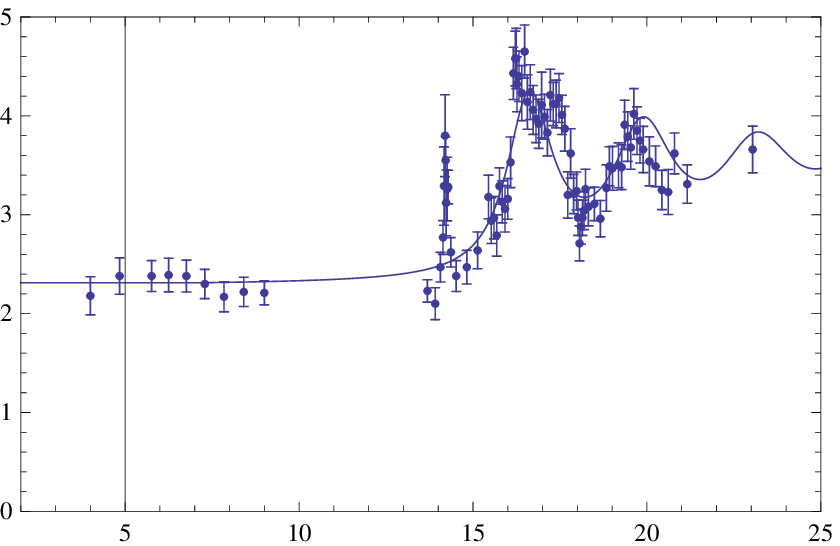} \qquad  
  \includegraphics[width=0.45\textwidth]{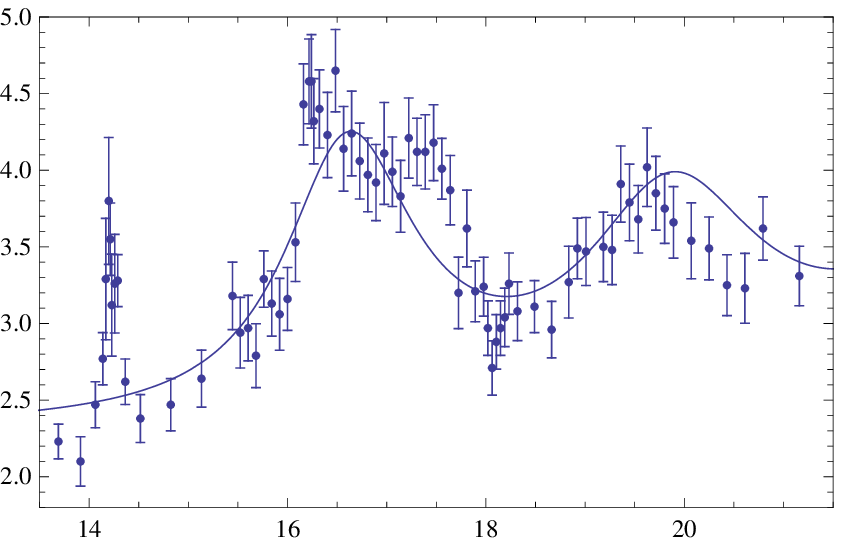}
 \end{center}
\caption{\label{fig:fit0} Simple fit to BES data for 
$R$ vs. $q^2/{\rm GeV}^2$. Right: Detailed View.}
\end{figure}

In order to obtain a more realistic picture of the $c\bar c$-- spectrum, 
we are going to fit the BES data 
\cite{Bai:2001ct,Ablikim:2006mb,Kuhn:2007vp} for the $R$--ratio in the 
$c\bar c$--region to a modified version of Shifman's model.
The spectra of $c\bar c$ mesons can be accounted for by
linear relations for the squared masses, $M^2_n=n \lambda^2 + M^2_0$,
$n=1, 2, 3,\ldots$, similarly to the case of the light 
mesons \cite{Gershtein:2006ng}. The trajectory of the $n^3S_1$ charmonia,
the $J^{PC}=1^{--}$ states
$\psi(3097)$, $\psi(3686)$, $\psi(4040)$, $\psi(4415)$, $\ldots$,
for instance, follows this pattern.  
Starting form the third resonance ($n=3$), these states can decay into open
charm and have widths of the order of $\Lambda_{QCD}$. The first two
are extremely narrow and may be described separately, but their properties
are unimportant for duality violation, which is related to the infinite tower
of high-$n$ resonances. We therefore choose an ansatz where the sum
over resonances begins at $n=3$ rather than $n=1$. A finite width
is included in analogy to (\ref{zdef}) and the variable $q^2$ is
shifted by a constant into $q^2-4 m^2_c$. This leads to the following
expression for the imaginary part of the correlator or, equivalently,
the $R$ ratio  
\begin{align}
 R &= R_{\rm light} - \frac{4}{3} \, \frac{1}{(1 - b/\pi) \, \pi} \,  
{\rm Im}\, \psi(3 + z) \,,
\qquad
 z= \left(-\frac{q^2 - 4 m_c^2 + i\epsilon}{\lambda^2} \right)^{1 - b/\pi} \,,
\label{step1}
\end{align}
The individual resonances are located at $q^2=n\lambda^2 +4 m^2_c$
($n=3,4,5,\ldots$) in the limit $b\to 0$. 
We observe that a rough description of the BES data
\cite{Bai:2001ct} can already be obtained with this formula,
where we find  $R_{\rm light} = 2.31$ from the measured R-ratio below charm 
threshold, $m_c=1.33~{\rm GeV}$,  $b \simeq 0.082$ and 
$\lambda^2 \simeq 3.08~{\rm GeV}^2$. This yields the result shown 
in Fig.~\ref{fig:fit0}, corresponding to a $\chi^2/{\rm d.o.f.} \simeq 2.5$.
We remark that our values for the fit parameters $m_c$ and $\lambda^2$
are in agreement with the results of \cite{Gershtein:2006ng}.
There is a second trajectory of $1^{--}$ charmonia, the $n^3D_1$ states.
Of these the first two resonances $\psi(3770)$ and $\psi(4160)$ are known.
The first one is barely above threshold and still rather narrow. It
may be considered separately, similar to $\psi(3097)$ and $\psi(3686)$.
Note also that $\psi(3770)$ is still below our default choice for the lower
limit of the high-$q^2$ region, $q^2 \geq 15\,{\rm GeV}^2$.
The remaining resonances $n^3D_1$ are rather close to the resonances 
$(n+1)^3S_1$ for $n\geq 2$. For an approximate treatment it appears
justified to subsume such a pair of close resonances under a single
peak and keep the ansatz given in (\ref{step1}). The accuracy of this 
description can be gauged by inspecting Fig.~\ref{fig:fit0}.
In any case, the normalization of the second term of $R$ in (\ref{step1})
is fixed in the large-$q^2$ limit by the free-quark result. 
%\begin{figure}[t]
%\begin{center}
%%\psfrag{x}[t]{$s$}
%%\psfrag{y}[b]{$(dB/ds)/B$}
%\resizebox{8cm}{!}{\includegraphics{fit0.eps}}
%\caption{\label{fig:fit0}
%the OPE approximation.}
%\end{center}
%\end{figure}

The fit could be refined by including the low-lying, narrow resonances
in the fit ansatz. Any finite number of resonances does not change
the asymptotic behaviour responsible for duality violation. Also the fit
parameters are not much affected by such modifications. For example, 
including the $\psi(3770)$ 
reduces the $\chi^2/{\rm d.o.f.} \to 1.7$, 
while the parameters of the continuum ansatz remain almost
unchanged, $R_{\rm light} = 2.26$,
$m_c=1.33~{\rm GeV}$,  $b \simeq 0.078$ and 
$\lambda^2 \simeq 3.08~{\rm GeV}^2$. 
We will therefore be content with the simple representation given
in (\ref{step1}) above.

%%%%%%%%%%%%%%%%%%%%%%%%%%%%%%%%%%%%%%%%%%%%%%%%%%%%%%%%%%%%%%%%%
%     Duality in B -> K l+l-                  
%%%%%%%%%%%%%%%%%%%%%%%%%%%%%%%%%%%%%%%%%%%%%%%%%%%%%%%%%%%%%%%%%
\section{\boldmath Quark-hadron duality in $B\to Kl^+l^-$}
\label{sec:dualitybk}

\subsection{General considerations}

The analytic structure of the matrix element of the operator
product in (\ref{khdef}) can be infered from
\begin{eqnarray}\label{khas}
-\frac{q^2}{8\pi^2}\langle {\cal K}^\mu_H\rangle &=&
i\int d^4x\, e^{iq\cdot x}\, 
\langle\bar K(k)|T\, j^\mu(x) H^c(0)|\bar B(p)\rangle\nonumber\\
&=&\sum_X
\frac{(2\pi)^3 \delta(\vec q+\vec k-\vec p_X)}{p_{X0}-q_0-k_0-i\epsilon}
\langle\bar K(k)|j^\mu(0)|X\rangle\langle X|H^c(0)|\bar B(p)\rangle\nonumber\\
&+&\sum_X
\frac{(2\pi)^3 \delta(\vec q+\vec p_X-\vec p)}{p_{X0}+q_0-p_0-i\epsilon}
\langle\bar K(k)|H^c(0)|X\rangle\langle X|j^\mu(0)|\bar B(p)\rangle
\end{eqnarray}

In order to discuss the properties of the matrix element in 
(\ref{khas}) we make the following simplifications.
First, we neglect the small penguin contributions in the weak
Hamiltonian, that is we take
\begin{equation}\label{hcnopeng}
H^c=C_1 Q^c_1 + C_2 Q^c_2
\end{equation}
Second, in the electromagnetic current we retain only the
charm-quark component, $j^\mu=Q_c \bar c\gamma^\mu c$, and
consider only contributions where the charm fields in $H^c$
are contracted with those in $j^\mu$. This neglects terms where
the charm-anticharm pair from $H^c$ annihilates into gluons
before connecting to the electromagnetic current. Such contributions
are of higher order in $\alpha_s$ and not essential for the
problem we want to address. 
 
We consider the matrix element in (\ref{khas}) as a function 
of $q_0$, keeping the kaon energy $k_0$ fixed at a value of order
$1\,\rm{GeV}$. This can be achieved by
injecting a spurion 4-momentum \cite{Chibisov:1996wf}
$r=(r_0,0,0,0)$ into the $H^c$ vertex. Then
$p+r=q+k$ and for $p_0$ and $k_0$ fixed at their physical values the 
variable $r_0=q_0+k_0-p_0$ grows with $q_0$. 
The physical kinematics is recovered for $r=0$.
Under the simplifying assumptions specified above, the
intermediate state $X$ in the first sum in (\ref{khas})
always contains a $c\bar c$ pair and a strange quark, in general
together with other hadronic states, and thus $p_{X0}-k_0\gsim 2m_c$. 
The state $X$ in the second sum contains a $c\bar c$ pair
and a $b$ quark, and $p_{X0}-p_0\gsim 2m_c$.
The matrix element in (\ref{khas}) is then seen to be an analytic
function of $q_0$ in the entire $q_0$-plane, except for two branch cuts 
at $2 m_c\lsim q_0 < \infty$ and at $-\infty < q_0\lsim -2 m_c$.
If we would relax the simplifications above and allowed
for intermediate states $X$ without $c\bar c$ pairs, the cuts
would extend down to lower values of $|q_0|$ on the real axis.

The OPE of the matrix element in (\ref{khas}) can be justified
for $q_0$ on the imaginary axis, sufficiently far from the origin,
that is at $q_0=i q_{0E}$, for $q_{0E} \gg\Lambda_{QCD}$.
The OPE defined in this way in the Euclidean can then
be analytically continued, term by term, from imaginary $q_0$
onto the positive real axis, corresponding to the Minkowskian domain.
Terms that are exponentially suppressed in $\Lambda_{QCD}/q_{0E}$
for large positive $q_{0E}$ become oscillating functions
of $q_0$ in the Minkowskian case, that is for large positive $q_0$
\cite{Chibisov:1996wf}. These oscillating terms are invisible at any 
finite order in the OPE and represent the duality violating contribution.

\subsection{Quantitative estimate of duality violation}

For a quantitative estimate of duality violation we have to resort
to a model of the hadronic correlator in (\ref{khas}).
To this end we write the Hamiltonian in (\ref{hcnopeng}) as
\begin{equation}\label{hca2}
H^c=a_2\, (\bar sb)_{V-A}(\bar cc)_{V-A}
\end{equation}
and assume a factorization of the currents, that is we neglect
interactions between $\bar cc$ and the $\bar B\to\bar K$ system.
The coefficient $a_2$ is then treated as a phenomenological parameter.
With these simplifications the correlator in (\ref{khas})
reduces to
\begin{equation}\label{khfact}
\langle {\cal K}^\mu_H\rangle = \frac{16\pi^2}{3} a_2\,
\langle (\bar sb)_{V-A}\rangle^\mu\, \Pi_c(q^2)
\end{equation}
where we omitted the longitudinal component $\sim q^\mu$.
$\Pi_c$ is the current correlator defined in (\ref{pimunu}).
The charm loop in (\ref{khfact}) contributes to the coefficient 
$a_9$ in the amplitude of $\bar B\to\bar Kl^+l^-$ a term
\begin {equation}\label{defd}
\Delta a_9 =a_2 d, \qquad\quad 
d\equiv\frac{16\pi^2}{3}\left(\Pi_c(q^2) - \Pi_c(0)\right)
\end{equation}
In the model of section \ref{sec:charmbes} we have
\begin {equation}\label{defpsi}
d=-\frac{4}{3}\frac{1}{1-b/\pi}\left[\psi(z+3)-\psi(z_0+3)\right]
\end{equation}
where
\begin{equation}
z=(-r-i\epsilon)^{1-b/\pi},\quad r=\frac{q^2-4m^2_c}{\lambda^2}\equiv
u(s-s_c),\quad u=m^2_B/\lambda^2
\end{equation}
and $z_0=z(q^2=0)$. For the parameters we use the following values
\begin{equation}\label{lambdamcb}
\lambda^2=3.08\,{\rm GeV}^2,\quad
m_c=1.33\,{\rm GeV},\quad b=0.082
\end{equation}
We will not employ the model (\ref{defd}) to describe the
entire charm-loop contribution, but only to estimate its
duality violating component. The remainder is more reliably
obtained by the OPE itself. In order to extract the term that
represents duality violation within the model, we decompose
\begin{equation}\label{psiz312}
\psi(z+3)-\psi(z_0+3)\equiv\left[\psi(-z-2)-\psi(z_0+3)-i\pi\right]_1
+\left[-\pi\cot \pi z + i\pi\right]_2
\end{equation}
This decomposition is useful for $q^2\gsim 15\,{\rm GeV}^2$, when
the second term starts being exponentially suppressed and gives the 
duality violating contribution. In correspondence with (\ref{psiz312})
we have $d\equiv d_1+d_2$ and the duality violating term is
\begin{equation}\label{defd2}
d_2=-\frac{4}{3}\frac{1}{1-b/\pi}\left[-\pi\cot \pi z + i\pi\right]_2
\approx-\frac{8\pi}{3}\exp(-2\pi br)\left(\sin 2\pi r - i\cos 2\pi r\right)
\end{equation}
When integrating the $|a_9|^2$ part of the $\bar B\to\bar Kl^+l^-$
rate over the high-$q^2$ region, from a lower limit $q^2_0=s_0 m^2_B$
to the end of the spectrum, the relative size of the duality violating
effect is then given by
\begin{equation}\label{rdv1def}
R_{DV,1}=\frac{2 a_2}{a_9} 
\frac{\int_{s_0}^{s_m}ds\,\lambda^{3/2}_K(s) f^2_+(s)\,{\rm Re}\, d_2}{
\int_{s_0}^{s_m}ds\,\lambda^{3/2}_K(s) f^2_+(s)}
\end{equation}
to first order in the charm-loop contribution. 
Here we have assumed $a_2$ to be real. Using the approximation
in (\ref{defd2}) and proceeding as in (\ref{dvrate}) we find
\begin{equation}\label{rdv1approx}
|R_{DV,1}|\lsim\frac{8}{3}\frac{a_2}{a_9}
\frac{\lambda^{3/2}_K(s_0) f^2_+(s_0)}{
\int_{s_0}^{s_m}ds\,\lambda^{3/2}_K(s) f^2_+(s)}\frac{1}{u}
\exp(-2\pi b u(s_0-s_c))
\end{equation}
This formula gives an excellent approximation of the full
result based on (\ref{rdv1def}). 
We note the (mild) exponential suppression and the power
suppression by $1/u=\lambda^2/m^2_B$. 
For $a_2$ we take the value $a_2=0.3$,
which is large enough to reproduce the measured $B\to K\psi$
branching fraction within the factorization ansatz. 
We recall that $a_9\approx 4$.   
Then, for $0.5 < s_0 < 0.6$, that is for $q^2_0$ in the vicinity
of $15\,{\rm GeV}^2$, the relative correction $|R_{DV,1}|$ 
is below $3\%$. In the rate for $\bar B\to\bar Kl^+l^-$,
$|a_{10}|^2$ is added to $|a_9|^2$, which roughly doubles the result.
The net effect of the uncertainty from (\ref{rdv1approx})
for the rate integrated over the high-$q^2$ region is therefore
only about $1.5\%$.    

As discussed in section \ref{sec:dualitytoy}, the second
order effect in $d_2$ is qualitatively different. It has
no cancellations due to oscillating terms, giving a positive
correction, and its impact increases with decreasing $b$.
If local duality is at least roughly fulfilled, as is the case
for high enough $q^2\gsim 15\,{\rm GeV}^2$, the duality violation
from $|d_2|^2$ is still suppressed, being of second order in the small 
quantity $a_2/a_9$. The relative size of this component is 
\begin{eqnarray}\label{rdv2}                                          
R_{DV,2} &=& \frac{a^2_2}{a^2_9}                                          
\frac{\int_{s_0}^{s_m}ds\,\lambda^{3/2}_K(s) f^2_+(s)\, |d_2|^2}{       
\int_{s_0}^{s_m}ds\,\lambda^{3/2}_K(s) f^2_+(s)}  \nonumber \\
&\approx&
\left(\frac{8\pi a_2}{3 a_9}\right)^2
\frac{\int_{s_0}^{s_m}ds\,\lambda^{3/2}_K(s) f^2_+(s)\, 
\exp(-4\pi b u(s-s_c))}{\int_{s_0}^{s_m}ds\,\lambda^{3/2}_K(s) f^2_+(s)} 
\end{eqnarray}
where the second term uses the approximation in (\ref{defd2}).
The approximate form and the full result in (\ref{rdv2}) agree
reasonably well. With our set of parameters we find
$R_{DV,2}=0.015$ at $q^2_0=15\,\rm{GeV}^2$, 
about half the size of $|R_{DV,1}|$.

%%%%%%%%%%%%%%%%%%%%%%%%%%%%%%%%%%%%%%%%%%%%%%%%%%%%%%%%%%
The phenomenological value $a_2=0.3$ used above 
is larger than the perturbative result for the coefficient $a_2$.
Thus it effectively absorbs the (partly unknown) effects
from factorizable and non-factorizable corrections.
We should emphasize here that the analytic structure of the latter is, 
in general, more complicated than it is implied by the approximation 
to $\langle{\cal K}_H^\mu\rangle$ in (\ref{khfact}).
Globally using the larger value of $a_2$ in our numerical estimate of
duality violation thus corresponds to the pessimistic scenario 
where the oscillating terms
from the non-factorizable corrections to the charm-loop are added
coherently. In reality,
we expect that at least some destructive interference between the
various contributions
appears, and the amount of duality violation should even be smaller than
our estimate.
%%%%%%%%%%%%%%%%%%%%%%%%%%%%%%%%%%%%%%%%%%%%%%%%%%%%%%%%%%

Our model for the charm-loop in (\ref{khfact}) is similar to
the ansatz originally proposed in \cite{Kruger:1996cv} and used
since then in many phenomenological studies. However, our motivation
for considering this model is essentially different. 
In contrast to \cite{Kruger:1996cv} it is not our goal to
model the hadronic effects on the spectrum point by point
in the $q^2$ distribution. More relevant than the detailed shape of
the spectrum is the rate integrated over the entire high-$q^2$ region,
which is best described in a model-independent way by the OPE, as
mentioned above. We rather employ the model to get an indication  
of the duality violating effects, which are not captured in an OPE
calculation. For this purpose the Shifman model for the current correlator
is adapted to the charm-quark case $\Pi_c$, with a choice of parameters
consistent with the basic features of the most recent experimental data
from BES. 
Note that the Kr\"uger-Sehgal ansatz for $\Pi_c$ \cite{Kruger:1996cv},
consisting of a spectral function with a finite number of resonances and 
a flat continuum for large $q^2$, contains no information on duality 
violation. The model we use to estimate duality violation
is very simple and involves many assumptions. Still we expect it to
indicate the systematics and the typical size of the effect, which
presumably is closely connected with the $c\bar c$ resonance structure
in the charm-loop contribution.

%%%%%%%%%%%%%%%%%%%%%%%%%%%%%%%%%%%%%%%%%%%%%%%%%%%%%%%%%%%%%%%%%
%     Comments
%%%%%%%%%%%%%%%%%%%%%%%%%%%%%%%%%%%%%%%%%%%%%%%%%%%%%%%%%%%%%%%%%
\section{Comments on the literature}
\label{sec:comments}

The high-$q^2$ region of $B\to K^*l^+l^-$ has also been analyzed
in the framework of an OPE in \cite{Grinstein:2004vb} and a recent
application was presented in \cite{Bobeth:2010wg}.
Our approach differs from the analysis of \cite{Grinstein:2004vb} in 
several respects. We go beyond the work of \cite{Grinstein:2004vb}
by addressing in detail the issue of duality violation and the
basis of the OPE formalism. In addition, we extend the OPE to include
second-order power corrections, which we estimate quantitatively.
A list of new results is given in the Conclusions. Here we would like to 
comment further on two conceptual differences between \cite{Grinstein:2004vb}
and our formulation. This concerns, first, the construction of operators
in the OPE and, second, the treatment of charm quarks.

The operators in our approach are built from $b$-quark fields in
full QCD rather than using HQET. This is convenient because the
operator basis is simpler and the OPE becomes particularly transparent.
Another advantage is that the matrix elements of the leading operators
are given by the usual form factors in full QCD. Unlike in a HQET
framework, not all dependence on $m_b$ is made explicit, but this is
not essential from a practical point of view and still allows the 
consistent inclusion of power corrections to a given order in the
expansion. This approach has been used for instance in the OPE for
inclusive $B$ decays applied to the computation of the lifetime
difference of $B_s$ mesons \cite{Beneke:1996gn,Beneke:1998sy}.
These investigations included power corrections \cite{Beneke:1996gn}
as well as corrections of ${\cal O}(\alpha_s)$ \cite{Beneke:1998sy}.
By contrast, in \cite{Grinstein:2004vb} a matching onto HQET operators
is performed from the start. This leads to a proliferation of
operators, whose matrix elements are not given by the usual form
factors. In fact, to reach a simplification of the resulting HQET
expressions, the authors of \cite{Grinstein:2004vb} partly undo the
matching , from HQET back to full QCD expressions.
We prefer to employ the full-QCD formulation throughout, in view
of the advantages mentioned above.

We next turn to the second point, the treatment of charm.
The authors of \cite{Grinstein:2004vb} perform an explicit
expansion in $m_c/m_b$, which corresponds to assuming the
hierarchy $m_c\ll m_b$. This implies that the charm quark is
not integrated out at the scale $m_b\approx\sqrt{q^2}$ and
continues to be an active field below this scale.
Operators with charm-quark fields are then present in the
OPE. 

We will argue that it is conceptually simpler to
integrate out charm immediately at the scale $m_b$,
assuming a hierarchy $m_b\sim m_c\gg\Lambda_{QCD}$,
and that this can be done without any loss in accuracy.
Charm-quark effects are then entirely contained in the coefficients
of the operators, as is apparent from the formulation of the OPE
given in section 3. 

To illustrate the point we consider the following example.
Let us take the scenario where $m_c\ll m_b$. In this case
the operator coefficients should be evaluated with $m_c=0$. 
On the other hand, additional operators involving
charm would have to be included in the OPE. 
With explicit $c\bar c$ fields, such operators arise at
dimension 6 or higher. For instance, radiating the virtual photon
from the two charm lines in the effective Hamiltonian (\ref{hcnopeng})
and leaving all four quark lines open, gives the dimension-6 operator
\begin{equation}\label{kh6c}                      
{\cal K}^\mu_{H6c} = 16\pi^2 Q_c\frac{q_\lambda}{q^4}\bigg[     
\bar c_i\gamma^\mu(1-\gamma_5)b_j\, \bar s_k\gamma^\lambda(1-\gamma_5)c_l  
-\{ \mu\leftrightarrow\lambda\}\bigg]   
\, (\delta_{ij}\delta_{kl} C_1 + \delta_{il}\delta_{kj} C_2)  
\end{equation}
in analogy to (\ref{kh6}). This operator has  $\bar B\to\bar K^{(*)}$
matrix elements contributing to the decay amplitude. Assuming next
that $m_c\gg\Lambda_{QCD}$, the contribution (\ref{kh6c}) can be
simplified by integrating out the charm fields in a further step.  
Contracting the charm lines into a loop, to which a gluon field
is attached, induces below $m_c$ the dimension-6 operator  
\begin{equation}\label{kh6d}                                                
{\cal K}^\mu_{H6d} = \frac{4}{3}C_1 Q_c\ln\frac{\mu^2}{m^2_c}
\, \frac{q_\lambda}{q^4}          
i\varepsilon^{\lambda\mu\alpha\beta}\,
\bar s\gamma_\alpha(1-\gamma_5)gD^\nu G_{\nu\beta}b
\end{equation}
The coefficient of this operator reproduces the logarithmic
$m_c$-dependence at this order in the OPE.
A similar discussion has been presented in \cite{Grinstein:2004vb},
illustrating how the logarithmic term $m^4_c\ln m^2_c$ in the
coefficient of the leading dimension-3 operator (\ref{kh3})
is recovered in an effective theory with four-quark operators
of the type $(\bar sb)(\bar cc)$. 
Whereas the latter effect vanishes for $m_c\to 0$, we note
that the coefficient in (\ref{kh6d}) has a logarithmic divergence 
in this limit. This is of no consequence if $m_c\gg\Lambda_{QCD}$,
where $m_c$ still represents a hard scale. In fact, for higher-dimensional
operators generated by (\ref{kh6c}),
that is operators with more factors of the gluon field and its derivatives,
the coefficients will scale as inverse powers of $m_c$.
A similar situation exists for inclusive semi-leptonic $b\to c$
decays, where four-quark operators with charm, analogous to (\ref{kh6c}),
also appear at third order in the OPE for $m_c\ll m_b$.
The effect of the corresponding charm loops has been refered to
as 'intrinsic charm' in \cite{Bigi:2005bh,Breidenbach:2008ua,Bigi:2009ym},
where the issue was discussed in great detail.
The above consideration indicates how the nonperturbative 
$\bar B\to\bar K^*$ matrix element of operators such as (\ref{kh6c})
can be treated in an expansion in $\Lambda_{QCD}/m_c$.
However, as demonstrated in \cite{Bigi:2005bh,Breidenbach:2008ua,Bigi:2009ym}
for semi-leptonic $b\to c$ decays, the same effects are also described
in a framework where charm is integrated out at the $m_b$ scale.
No active charm fields need to be considered in this case and
(\ref{kh6c}) is absent from the OPE. 
A difference between the OPE with or without active charm
is that the framework with charm fields and with a strong hierachy
$m_c\ll m_b$ assumed, would offer the possibility of efficiently
resumming
logarithmic terms $\ln m_b/m_c$. Since such logarithms are not very
large and in view of the additional power suppression of such terms,
such resummations appear not to be necessary in practice.
We also stress that effects such as (\ref{kh6c}), irrespective of their
detailed treatment, are suppressed at least as $1/m^3_b$.
Small differences in the method of their calculation are therefore hardly
relevant. 
We thus conclude that integrating out charm at the scale $m_b$ 
is entirely justified. The OPE is then constructed in a single step 
and with a simpler operator basis. For these reasons
the approach appears preferable and we have adopted it here.

%%%%%%%%%%%%%%%%%%%%%%%%%%%%%%%%%%%%%%%%%%%%%%%%%%%%%%%%%%%%%%%%%
%     Conclusions
%%%%%%%%%%%%%%%%%%%%%%%%%%%%%%%%%%%%%%%%%%%%%%%%%%%%%%%%%%%%%%%%%
\section{Conclusions}
\label{sec:conclusion}

The amplitude for $\bar B\to\bar Ml^+l^-$, 
$\bar M=\bar K, \bar K^*,\ldots$, has the general form given in (\ref{abmll}).
It contains the component
\begin{equation}
A^\mu_9 = C_9\, \langle\bar M|\bar s\gamma^\mu(1-\gamma_5)b|\bar B\rangle     
+\, C_7\, \frac{2i m_b}{q^2} q_\lambda\,
\langle\bar M|\bar s\sigma^{\lambda\mu}(1+\gamma_5)b|\bar B\rangle
+ \langle\bar M|{\cal K}^\mu_H(q)|\bar B\rangle 
\end{equation}
\begin{equation}\label{khconcl}          
{\cal K}^\mu_H(q)=
-\frac{8\pi^2}{q^2} i\int d^4x\, e^{iq\cdot x}\, T\, j^\mu(x) H^c(0)
\end{equation}
which receives a nonlocal, hadronic contribution 
$\langle{\cal K}^\mu_H\rangle\equiv
\langle\bar M|{\cal K}^\mu_H|\bar B\rangle$ from the 
matrix element of the nonleptonic weak Hamiltonian $H^c$ in addition to the
semileptonic form factor terms ($\sim C_9,C_7$). Although the hadronic part
$\langle {\cal K}^\mu_H\rangle$ is relatively small numerically
outside the narrow-resonance region ($\sim 10\%$ of $A_9$), it needs to 
be reliably computed in order to achieve very accurate predictions. 
We have presented a detailed study of the hadronic contribution in the region
of large dilepton invariant mass $q^2\gsim 15\,{\rm GeV}^2$, based on an
operator product expansion in inverse powers of the hard scale $\sqrt{q^2}$.
Working with $b$-quark fields in full QCD and factorizing the dependence
on $m_b$, $\sqrt{q^2}$ and $m_c$ into the coefficient functions, we obtain
the following results:
\begin{itemize}
\item
To leading order in the OPE and to all orders in $\alpha_s$,
$\langle{\cal K}^\mu_H\rangle$ is expressed in terms of the standard form 
factors parametrizing the matrix elements
$\langle\bar s\gamma^\mu(1-\gamma_5)b\rangle$ and
$\langle\bar s\sigma^{\lambda\mu}(1+\gamma_5)b\rangle$
of dimension-3 operators, up to coefficients calculable in perturbation
theory. To lowest order in $\alpha_s$ only 
$\langle\bar s\gamma^\mu(1-\gamma_5)b\rangle$ is present.
\item
In the chiral limit ($m_s=0$), the first power corrections appear
only at second order ($\sim 1/q^2$) and are governed by dimension-5
operators with a gluon field of the form $\bar sgG\Gamma b$. The corrections 
are computed explicitly, using the limit $E_K\gg\Lambda_{QCD}$ for the 
hadronic matrix elements, and are shown to be smaller than $1\%$ for $A_9$.
\item
For $m_s\not= 0$ the dimension-4 operators 
$m_s\bar s\gamma^\mu(1+\gamma_5)b$ and
$m_s\bar s\sigma^{\lambda\mu}(1-\gamma_5)b$, with right-handed strange
quarks, can arise in the OPE. Because they are absent at order $\alpha_s^0$,
their contribution will be suppressed to the negligible level of
$\alpha_s m_s/m_b\sim 0.5\%$. 
Besides, no new form factors will be introduced by these operators. 
\item
Within the OPE framework the effect of weak annihilation is a natural 
ingredient, which we have briefly discussed, mainly for illustration.
Since in addition to being a third-order power correction it comes
with small Wilson coefficients, its numerical impact of a few permille
is entirely negligible. 
\item
We have clarified the relationship between the OPE at high $q^2$ and
QCD factorization at low $q^2$ by showing that both descriptions yield 
consistent results at intermediate values of $q^2\approx 15\,{\rm GeV}^2$.
\item
A relevant topic at high $q^2$ is the issue of quark-hadron duality,
which is closely related to the existence of an OPE. We have defined
the OPE with the help of a spurion momentum that allows for an independent 
scaling of $q^2$ at fixed $m_B$ and kaon energy.  This allows one to
clarify the analytic structure of the matrix elements of the correlator
(\ref{khconcl}). We then employed a model based on an infinite series
of charm resonances to estimate quantitatively the amount of
duality violation, resulting in about $\pm 2\%$ for the rate integrated
over the high-$q^2$ region. An important aspect is the different
sensitivity to duality violation of the contributions to the rate 
linear or quadratic in $\langle {\cal K}^\mu_H\rangle$. 
The quadratic term is more vulnerable to duality violation, but
numerically suppressed as a term of second order in the
small charm-loop contribution.    
The systematics of duality have further been studied in a
toy model for the rare decays, in which the factorization of the 
charm loop is exact.
\end{itemize}

The main conclusion is that the high-$q^2$
region of $B\to K^{(*)}l^+l^-$ is theoretically under excellent control.
Decay rates and distributions are perturbatively calculable up to the 
nonperturbative effects accounted for by the standard form factors in 
full QCD. Further nonperturbative corrections are strongly suppressed
and negligible within the accuracy of a few percent. 
The existence of the OPE
implies that at high $q^2$ the theory of $B\to K^{(*)}l^+l^-$ has an
even more solid basis than at low $q^2$.
An example for applications is the combined analysis of
$B\to Kl^+l^-$ and $B\to K\nu\bar\nu$ \cite{Bartsch:2009qp,Hewett:2004tv}. 
Here the form factor dependence can be essentially eliminated, which leads to 
precision observables sensitive to new physics effects \cite{Bartsch:2009qp}.
Beyond $B\to K^{(*)}l^+l^-$, the results of our analysis apply to the 
high-$q^2$ region of exclusive rare decays with similar final states such
as $B\to K\pi l^+l^-$, $B_s\to\phi l^+l^-$ and, with appropriate 
modifications, to $B\to \rho l^+l^-$ or $B\to\pi l^+l^-$.

%%%%%%%%%%%%%%%%%%%%%%%%%%%%%%%%%%%%%%%%%%%%%%%%%%%%%%%%%%%%%%%%%
%     Appendix
%%%%%%%%%%%%%%%%%%%%%%%%%%%%%%%%%%%%%%%%%%%%%%%%%%%%%%%%%%%%%%%%%
\appendix

\section{Complete basis of operators through dimension 4}

We show that the OPE of ${\cal K}^\mu_H$ in (\ref{khdef}) can be
expressed in terms of
\begin{equation}\label{o31o32}                                                
{\cal O}^\mu_{3,1} = \left(g^{\mu\nu}-\frac{q^\mu q^\nu}{q^2}\right)         
\, \bar s\gamma_\nu(1-\gamma_5)b\ \, ,  \qquad\quad                  
{\cal O}^\mu_{3,2} = \frac{i m_b}{q^2} q_\lambda\,                           
\bar s\sigma^{\lambda\mu}(1+\gamma_5)b                             
\end{equation}
at the level of dimension-3 and dimension-4 operators in full QCD,
in the chiral limit $m_s=0$. For $m_s\not= 0$, $m_s\ll m_b$, right-handed
strange quarks can also appear, together with a suppression factor of 
$m_s/m_b$. This leads to the two additional operators ${\cal O}^\mu_{4,1}$
and ${\cal O}^\mu_{4,2}$ in (\ref{o41}) and (\ref{o42}), which can be
counted as operators of dimension 4.
The matrix elements of ${\cal O}^\mu_{3,1}$, ${\cal O}^\mu_{3,2}$,
${\cal O}^\mu_{4,1}$ and ${\cal O}^\mu_{4,2}$ between $\bar B$ and
$\bar K^{(*)}$ are all given by the standard $\bar B\to\bar K^{(*)}$
form factors.  

Together this implies that at leading (dimension 3) and next-to-leading
order (dimension 4) in the OPE, and to all orders in $\alpha_s$, standard
form factors are the only hadronic matrix elements required. Corrections
to these terms arise only at second order ($\sim \Lambda^2/q^2$) in the OPE
through operators of dimension 5. In other words, there are no 
first-order power corrections ($\sim\Lambda/m_b$) in the OPE of
${\cal K}^\mu_H$, except for the purely kinematical dependence on
$q^2$ and $m^2_b$ in the coefficient functions. These functions contain
first-order terms such as $(m^2_b-q^2)/m^2_b$, which however are
calculable and do not introduce unknown hadronic matrix elements.

We demonstrate the completeness of the basis
${\cal O}^\mu_{3,1}$, ${\cal O}^\mu_{3,2}$, ${\cal O}^\mu_{4,1}$ and 
${\cal O}^\mu_{4,2}$ for operators of dimension 3 and 4 by enumerating
all possibilities consistent with the relevant symmetries,
and using the equations of motion.

We first assume $m_s=0$. Operators of dimension 3 built from
$\bar s_L$ and $b$ have the form
\begin{equation}\label{opdim3}
(\bar s_L\Gamma b)^\mu\, , \qquad\quad \Gamma=1,\, \gamma^\alpha,\,
\sigma^{\alpha\beta}
\end{equation}
where Lorentz indices can be contracted with the metric  $g$,
the $\varepsilon$-tensor and factors of $q$ to yield a 4-vector with 
index $\mu$. Useful relations are
\begin{equation}\label{qpartial}
q^\mu (\bar s_L\Gamma b) = i\partial^\mu (\bar s_L\Gamma b)
\end{equation}
which holds for $\bar B(p)\to \bar K^{(*)}(k)$ matrix elements, and
\begin{equation}\label{partialdd}                                          
\partial_\mu (\bar s_L\Gamma b)=
\bar s_L\Gamma \overleftarrow{D}_\mu b + \bar s_L\Gamma D_\mu b       
\end{equation}
or similar identities.
Exactly three structures can be formed:
\begin{equation}\label{opdim3b}
\bar s_L\gamma^\mu b\, ,\qquad
q_\nu \bar s_L\sigma^{\mu\nu} b\, ,\qquad
q^\mu \bar s_L  b
\end{equation}
This can be seen as follows. For $\Gamma=1$ in (\ref{opdim3}) the
only possibility is the third operator in (\ref{opdim3b}).
For $\Gamma=\gamma^\alpha$ we obtain the first operator in (\ref{opdim3b})
and
\begin{equation}
q^\mu q_\nu \bar s_L\gamma^\nu b=
q^\mu (\bar s_Li\not\!\!\overleftarrow{D}b+\bar s_L i\not\!\! D b)=
m_b q^\mu \bar s_L b
\end{equation}
which is equivalent to the third term in (\ref{opdim3b}).
For $\Gamma=\sigma^{\alpha\beta}$ we can write the second
operator in (\ref{opdim3b}). In addition, we could form
\begin{equation}                                                               
\varepsilon^{\mu\alpha\lambda\nu} q_\alpha
\bar s_L \sigma_{\lambda\nu}b \sim
q_\alpha \bar s_L \sigma^{\mu\alpha}\gamma_5 b=
q_\alpha \bar s_L \sigma^{\mu\alpha}b
\end{equation}
which leads back to the same structure. This exhausts the possibilities
and proves the completeness of the basis in (\ref{opdim3b}) at the
dimension-3 level.

We next consider operators of dimension 4. These have, in addition to the
fields in (\ref{opdim3}), one covariant derivative acting on the 
strange quark. Their general form is
\begin{equation}\label{opdim4}                                            
(\bar s_L \overleftarrow{D}\Gamma b)^\mu 
\end{equation}
Because of
\begin{equation}
\bar s_L \overleftarrow{D}\Gamma D_\lambda b=
\partial_\lambda (\bar s_L \overleftarrow{D}\Gamma b) -
\bar s_L \overleftarrow{D}\overleftarrow{D}_\lambda\Gamma b=
-i q_\lambda (\bar s_L \overleftarrow{D}\Gamma b) + \rm{dim\ 5}
\end{equation}
terms with extra covariant derivatives acting on $b$ do not lead
to independent dimension-4 operators and may be disregarded.

For $\Gamma=1$ in (\ref{opdim4}) we can write the operator
$\bar s_L \overleftarrow{D}^\mu b$. Using the identity in (\ref{sdbq})
this operator can be expressed in terms of the basis in (\ref{opdim3b}): 
\begin{equation}\label{sdbop3}
\bar s_L \overleftarrow{D}^\mu b=\frac{i}{2}m_b \bar s_L\gamma^\mu b
-\frac{1}{2} q_\nu \bar s_L\sigma^{\mu\nu}b-\frac{i}{2}q^\mu \bar s_L b
\end{equation}
The other possible structure can be reduced as
\begin{equation}\label{oneqq}
q^\mu q_\nu\bar s_L \overleftarrow{D}^\nu b=
\frac{i}{2}(m^2_b-q^2) q^\mu \bar s_L b
\end{equation}

For $\Gamma=\gamma^\alpha$ the dimension-4 operator has the form
\begin{equation}\label{sdgamb}
\bar s_L\overleftarrow{D}^\nu\gamma^\alpha b
\end{equation}
We list the possible contractions of (\ref{sdgamb}) into a 4-vector
together with the reduction to the basis in (\ref{opdim3b}), neglecting
operators of dimension 5.
\begin{eqnarray}
q^\mu \bar s_L\not\!\!\overleftarrow{D} b &=& 0 \\
q^\mu q_\nu q_\alpha \bar s_L\overleftarrow{D}^\nu\gamma^\alpha b &=&
\frac{i}{2} m_b (m^2_b-q^2) q^\mu \bar s_L b \\
-i q_\nu \bar s_L\overleftarrow{D}^\nu\gamma^\mu b &=&
\bar s_L\overleftarrow{D}^\nu\gamma^\mu D_\nu b \nonumber\\
&=& \frac{1}{2}\partial^\nu\partial_\nu(\bar s_L\gamma^\mu b)-
\frac{1}{2}\bar s_L\gamma^\mu D^\nu D_\nu b =
\frac{m^2_b-q^2}{2}\bar s_L\gamma^\mu b \label{gammab}\\
q_\alpha \bar s_L\overleftarrow{D}^\mu\gamma^\alpha b &=&
\bar s_L\overleftarrow{D}^\mu i\not\!\! D b =
m_b \bar s_L\overleftarrow{D}^\mu b \label{gammac}\\
i\varepsilon^{\mu\alpha\lambda\nu} q_\alpha
\bar s_L\overleftarrow{D}_\lambda\gamma_\nu b &=&
q_\alpha \bar s_L\overleftarrow{D}_\lambda
(\gamma^\lambda\gamma^\mu\gamma^\alpha -g^{\lambda\mu}\gamma^\alpha
-g^{\alpha\mu}\gamma^\lambda + g^{\alpha\lambda}\gamma^\mu)b =\nonumber\\
&=&-q_\alpha \bar s_L\overleftarrow{D}^\mu\gamma^\alpha b
+ q^\lambda \bar s_L\overleftarrow{D}_\lambda\gamma^\mu b \label{gammad}
\end{eqnarray}
The r.h.s. of (\ref{gammac}) is reduced to the basic operators
via (\ref{sdbop3}). The terms on the r.h.s. of (\ref{gammad})
lead back to the expressions (\ref{gammac}) and (\ref{gammab}).

For $\Gamma=\sigma^{\alpha\beta}$ the dimension-4 operator becomes
\begin{equation}\label{sdsigb}
\bar s_L\overleftarrow{D}^\lambda\sigma^{\alpha\beta} b
\end{equation}
We list again the possible contractions and their reduction
to the operators in (\ref{opdim3b}).
\begin{eqnarray}
\bar s_L \overleftarrow{D}_\lambda\sigma^{\lambda\mu} b &=& 
i \bar s_L \overleftarrow{D}_\lambda
(\gamma^\lambda\gamma^\mu-g^{\lambda\mu}) b =
-i \bar s_L \overleftarrow{D}^\mu b \label{sigmaa}\\
q_\lambda q_\beta \bar s_L\overleftarrow{D}^\lambda\sigma^{\mu\beta} b
&=& i m_b q_\lambda \bar s_L\overleftarrow{D}^\lambda\gamma^\mu b
-i q_\lambda q^\mu \bar s_L\overleftarrow{D}^\lambda b \label{sigmab}\\
\varepsilon^{\mu\lambda\alpha\beta}
\bar s_L\overleftarrow{D}_\lambda\sigma_{\alpha\beta} b
&\sim & \bar s_L\overleftarrow{D}_\lambda\sigma^{\lambda\mu} b \label{sigmac}
 \end{eqnarray}
The r.h.s. of (\ref{sigmaa}) is given by (\ref{sdbop3}). The r.h.s. of 
(\ref{sigmab}) is equivalent to (\ref{gammab}) and (\ref{oneqq}),
and (\ref{sigmac}) reduces to (\ref{sigmaa}).

This completes the proof that all operators of dimension 3 and 4,
(\ref{opdim3}) and (\ref{opdim4}), can be expressed in terms of 
(\ref{opdim3b}) for $m_s=0$. 
Taking current conservation into account
leaves us with ${\cal O}^\mu_{3,1}$ and ${\cal O}^\mu_{3,2}$ in 
(\ref{o31o32}). Similar arguments hold if $\bar s_L$ is replaced by
$\bar s_R$. Since right-handed strange quarks come with a factor $m_s$,
the two further operators ${\cal O}^\mu_{4,1}$ and 
${\cal O}^\mu_{4,2}$ in (\ref{o41}) and (\ref{o42}) are obtained
for $m_s\not= 0$.   

Similarly to the derivations above it can be shown that 
dimension-5 operators of the form 
$\bar s_L \overleftarrow{D}_\alpha \overleftarrow{D}_\beta \Gamma b$,
contracted to a Lorentz vector with index $\mu$, can always be reduced
to linear combinations of (\ref{opdim3b}) and operators containing
a factor of the gluon field strength $G_{\alpha\beta}$.
The former terms correspond to dimension-3 operators with a purely
kinematic power suppression. Genuine operators of dimension 5 therefore
have the form $(\bar s_L G\Gamma b)^\mu$, as quoted in (\ref{o5n}).

\section{Numerical input}

In this appendix we collect input we have used in 
numerical calculations.
The numbers quoted are our central values.
The form factors have an uncertainty of roughly $\pm 15\%$.

The $B\to K^*$ form factors are parametrized as \cite{Ball:2004rg}
\begin{equation}\label{a0q2par}                                                      
A_0(q^2)=\frac{1.364}{1-q^2/(5.28\,\rm{GeV})^2}-                                   
\frac{0.990}{1-q^2/(36.78\,\rm{GeV}^2)}                                          
\end{equation}
\begin{equation}\label{vq2par}                                                   
V(q^2)=\frac{0.923}{1-q^2/(5.32\,\rm{GeV})^2}-                                    
\frac{0.511}{1-q^2/(49.40\,\rm{GeV}^2)}                                           
\end{equation}

The $B\to K$ form factor is parametrized as \cite{Bartsch:2009qp}
\begin{equation}\label{fpkpar}
f_+(s)=f_+(0)\frac{1-(b_0+b_1-a_0 b_0)s}{(1-b_0 s)(1-b_1 s)}\, ,
\qquad\quad s\equiv\frac{q^2}{m^2_B}\, , 
\qquad b_0\equiv \frac{m^2_B}{m^2_{B^*_s}}
\end{equation}
with the default choice $f_+(0)=0.304$, $a_0$=1.6 and $b_1=b_0$.

Further parameters are summarized in Table~\ref{tab:parinput}.
\begin{table}[t]
\centering
\begin{tabular}{|c|c|c|c|}
  \hline\hline
   $m_K$ & $f_K$ &
   $f_\parallel$ \cite{Beneke:2003zv} & 
   $f_\perp$ \cite{Ball:2006eu} \\
  \hline
   $0.496$ & $0.16$  & $0.218$  & $0.185$ \\
  \hline\hline
   $m_B$ & $m_{B^*_s}$ &
   $f_B$ & $\lambda_B$ \\
  \hline
   $5.28$ & $5.41$  & $0.2$  & $0.350$ \\
  \hline\hline
   $\bar m_b$ & $\bar m_c$ & 
   $\Lambda_{\overline{\rm MS},5}$ & $m_{K^*}$ \\
  \hline
   $4.2$ & $1.3$ & $0.225$ &  $0.894$ \\
  \hline\hline
%$G_F^2 \big[{\rm GeV}^{-5}\,{\rm ps}^{-1}\big]$  $206.52$
\end{tabular}
\caption{\label{tab:parinput} Input parameters in ${\rm GeV}$. 
$f_\perp=f_\perp(1\,\rm{GeV})$.}
\end{table}

%%%%%%%%%%%%%%%%%%%%%%%%%%%%%%%%%%%%%%%%%%%%%%%%%%%%%%%%%%%%%%%%%
%     Acknowledgements
%%%%%%%%%%%%%%%%%%%%%%%%%%%%%%%%%%%%%%%%%%%%%%%%%%%%%%%%%%%%%%%%%
\section*{Acknowledgements}
We thank the Galileo Galilei Institute for Theoretical Physics for
hospitality and the INFN for partial support during the initial
phase of this project. G.B. thanks the CERN Theory Division
for support and hospitality while this paper was being completed.
We thank Alexander Khodjamirian and Thomas Mannel for discussions
and critical remarks.
This work was supported in part by the DFG cluster of excellence
`Origin and Structure of the Universe' and by the DFG
Graduiertenkolleg GK 1054.

%%%%%%%%%%%%%%%%%%%%%%%%%%%%%%%%%%%%%%%%%%%%%%%%%%%%%%%%%%%%%%%%%
%     References
%%%%%%%%%%%%%%%%%%%%%%%%%%%%%%%%%%%%%%%%%%%%%%%%%%%%%%%%%%%%%%%%%

\end{document}